\documentclass[12pt]{article}
\usepackage{epsfig}
\def\Tr{{\rm Tr}\, }
\def\be{\begin{eqnarray}}
\def\ee{\end{eqnarray}}
\def\lb{\label}\def\appendix{{\newpage\section*{Appendix}}\let\appendix\section%
        {\setcounter{section}{0}
        \gdef\thesection{\Alph{section}}}\section}

\begin{document}
\hfill{\tt ITFA-2003-11}\\\mbox{}
\hfill {\tt hep-th/0303006}\\
\vskip0.3truecm
\begin{center}
\vskip 2truecm {\Large\bf A holographic reduction of Minkowski
space-time  
}\\
\vskip 1truecm
{\large\bf
Jan de Boer${}^1$\footnote{\tt jdeboer@science.uva.nl}  and
Sergey N.~Solodukhin${}^2$\footnote{
{\tt soloduk@theorie.physik.uni-muenchen.de}}
}\\
\vskip 0.6truecm
\it{$^1$Instituut voor Theoretische Fysica,   \par
Valckenierstraat 65, 1018XE Amsterdam, The Netherlands}\\
\vskip 0.4truecm
\it{
${}^2$Theoretische Physik,
Ludwig-Maximilians Universit\"{a}t,\par
Theresienstrasse 37,
D-80333, M\"{u}nchen, Germany}
\vskip 1truemm
\end{center}
\vskip 1truecm
\begin{abstract}
\noindent

Minkowski space can be sliced, outside the lightcone, in terms of
Euclidean Anti-de Sitter and Lorentzian de Sitter slices. In this
paper we investigate what happens when we apply holography to each
slice separately. This yields a dual description living on two
spheres, which can be interpreted as the boundary of the light
cone. The infinite number of slices gives rise to a continuum
family of operators on the two spheres for each separate bulk
field.
For a free field we explain how the Green's function and
(trivial) S-matrix in Minkowski space can be reconstructed in terms of
two-point functions of some putative conformal field theory on the
two spheres. Based on this we propose a Minkowski/CFT correspondence
which can also be applied to interacting fields.
We comment on the interpretation of the
conformal symmetry of the CFT, and on generalizations to curved
space.

\end{abstract}
\vskip 1cm
\newpage

\section{Introduction}
\setcounter{equation}0

Holography is a powerful principle, that after it had been
proposed in \cite{tHooft} and \cite{Susskind} has found a concrete
and quantitative realization in string theory in terms of the
AdS/CFT correspondence \cite{Malda,Gubs, Wit}. Since then, many
generalizations have been found, but all of these are deformations
of, or limits of, the AdS/CFT correspondence. There have been
attempts to generalize the AdS/CFT correspondence to de Sitter
space (this was first discussed in \cite{Wit1} and \cite{Str}) but
unfortunately string theory on de Sitter space, the dual field
theory and the precise form of the correspondence are not well
understood.

Besides Anti-de Sitter space and de Sitter space, which are
solutions of the Einstein equations with negative and positive
cosmological constant, there is a third class of spaces, namely
those with zero cosmological constant. In particular, Minkowski
space belongs to this class, and one may wonder what one can say
about the existence and nature of a holographic dual description.
There are local versions of entropy bounds \cite{enbound} that
also apply to Minkowski space, but these provide no specific
insight about any dual description. An alternative approach is to
take the decompactification limit of the AdS/CFT correspondence
\cite{polchinski}. Though this is in principle a viable procedure,
it has not led to an explicit dual description.

In this paper we will try something different. Minkowski space
with metric
\begin{equation} \label{eq1}
ds^2 = -dX_0^2 + dX_1^2 + \ldots + dX_{d+1}^2
\end{equation}
has three regions depending on whether $t^2 <0$, $t^2=0$, or
$t^2>0$, with
\begin{equation} \label{eq2}
-t^2 \equiv X\cdot X =-X_0^2 + X_1^2 + \ldots + X_{d+1}^2 .
\end{equation}
The regions with $t^2>0$ and $t^2<0$ correspond to the inside and
the outside of the light-cone respectively, whereas $t^2=0$ is the
light-cone itself. For a fixed value of $t^2$, equation
(\ref{eq2}) describes a slice of Minkowski space whose geometry is
either Euclidean anti-de Sitter, flat or Lorentzian de Sitter. The
idea will be to apply holography to each slice separately, and in
this way to effectively reduce the number of dimensions by one.
All slices asymptote to the past or future light-cone, and
therefore all dual field theories live on two d-spheres, which can
be thought of as the boundaries of the past and future light-cone.
It would appear as if we have lost two dimensions instead of one,
but because of the infinite number of slices, we get an infinite
number of fields on the two spheres, and this effectively
reinstates one of the two lost dimensions. However, we cannot
simply interpret the infinite number of fields as coming from a KK
reduction of a theory in one dimension higher. The asymptotics of
the metric (\ref{eq1}) suggest that the extra dimension is a null
dimension. Instead of trying to make sense of theories living on a
space with a degenerate metric, we will mostly work directly with the
infinite set of fields itself.

There are some obvious problems with this procedure. First of all,
Minkowski space has de Sitter slices, and a holographic reduction
along those slices requires us to apply the ill-understood dS/CFT
correspondence. Luckily, for the purposes in this paper, we will
only need some basic information like the form of two-point
functions, and these are almost completely fixed by the symmetries
of de Sitter space. Ultimately, if this whole procedure turns out
to be consistent, one may try to reverse the logic and try to use
known properties of Minkowski space in order to study holography
for de Sitter space, but that is beyond the scope of the present
paper.

Another problem is the presence of the light-cone itself. Only
non-interacting massless fields can spend their entire life inside
the light-cone, but there are few if any of such degrees of
freedom so it seems completely safe to leave the light-cone itself
outside the discussion.

Despite these problems, we will find some encouraging results. For
a free field we explain how the Green's function and S-matrix in
Minkowski space can be reconstructed in terms of two-point
functions of some putative conformal field theory on the two
spheres. That this is at all possible is related to the following
fact. Consider two points $X$ and $Y$ on Minkowski space, with
$-t^2=X\cdot X$ and $-u^2=Y\cdot Y$. Of crucial importance is the
distance $d$ between $X$ and $Y$, which is given by
$$ d^2 = (X-Y) \cdot (X-Y). $$ For simplicity we assume that $t^2$
and $u^2$ have the same sign. Then $\tilde{X}=X/|t|$ and
$\tilde{Y}=Y/|u|$ are coordinates on the same (anti)de Sitter
space. Clearly,
$$ d^2 = -u^2 - t^2 - 2 |ut| \tilde{X} \cdot \tilde{Y}. $$
Now the geodesic distance on (anti)de Sitter space is a function
of $\tilde{X} \cdot \tilde{Y}$ only, and we see that the geodesic
distance in Minkowski space is a function of $t$, $u$ and the
geodesic distance in (anti)de Sitter space only. This fact is
essential in relating Minkowski space quantities to correlation
functions in the CFT's.

The emerging picture is rather intriguing. It suggests that it must
be possible to completely re-formulate the  Quantum Field
Theory in Minkowski space entirely in terms of correlation
functions of a set of conformal operators associated with the past and future
of the light-cone. We consider this as
a step towards a better understanding of the holography
(well established in anti-de Sitter spacetime) in asymptotically flat
space-times.

An important element in formulation of a holographic duality is the
identification of the underlying conformal symmetry.
In asymptotically anti-de Sitter spaces
this symmetry arises as asymptotic symmetry
near the time-like infinity \cite{Brown}.  The class of asymptotically flat
spaces and the respective symmetries near the null-infinity
(where the holographic dual could naively be thought to live) were
studied long ago in \cite{Bondi:1962px} (see also \cite{Sommers}
and \cite{Ashtekar} for a related discussion of the space-like infinity). 
The group of
symmetries is the so-called BMS group, an infinite dimensional
abelian group, which does not contain the relevant conformal
group. The idea pushed forward in this paper is to look at the
infinity of the light-cone and uncover the asymptotic symmetries
there. The allowed set of spaces is larger than the set of
asymptotically flat spaces, and the group of asymptotic symmetries
contains the BMS group but also the conformal group. The
appearance of the conformal symmetry in this case is quite
transparent. It is just an extension of the well-studied conformal
symmetry near the boundary of a anti-de Sitter slice to a symmetry
near the boundary of the light-cone in Minkowski space. The price
to pay for all this is that translation invariance of Minkowski
space is not manifest, but takes a much more complicated form.

The outline of this paper is as follows. In section~2 we discuss
in more detail the slicing of Minkowski space-time and its
properties. In section~3 we consider free massless scalar fields,
and their reduction on each of the slices. In section~4 we
consider the finite part of the on-shell action for a free scalar
field, and will see that the structure suggests the presence of
conformal correlation functions in Minkowski space quantities. In
section~5 we will make the relation between correlation functions
in the conformal field theory and bulk quantities more precise,
and we use this to reconstruct the Minkowski space S-matrix and
Green's function. In section~6 we look at more general spaces. We
introduce an appropriate notion of asymptotically Minkowski space,
analyze the symmetries that preserve these asymptotics. We also
examine the way conformal symmetries are implemented and whether
it is meaningful to view the infinite set of conformal operators
as fields living in a space with one additional dimension, and
comment on the meaning of translation invariance and the BMS
group. In section~7 we study gravity in Minkowski space, and how
the space-time metric is encoded in data on the two spheres.
Finally, we present some conclusions and further speculations.

\section{Slicing Minkowski space}
\setcounter{equation}0 In this section we consider the foliation
of Minkowski spacetime with (positive or negative) constant
curvature subspaces that is naturally associated to the light-cone
structure of the space-time. Let us consider ($d+2$)-dimensional
Minkowski space with coordinates $X_0,X_1,...,X_{d+1}$ and metric
\be
ds^2=-dX_0^2+dX_1^2+...+dX^2_{d+1}~~.
\lb{1}
\ee
The light-cone defined as

$
\noindent  {\cal C}:~~-X_0^2+X^2_1+...+X^2_{d+1}=0
$

\noindent naturally splits the space-time in three  regions

\medskip

\noindent 1.  ${\cal D}:~~-X_0^2+X^2_1+...+X^2_{d+1}>0$, the region
lying outside  the light-cone ${\cal C}$;

\medskip

\noindent 2.   ${\cal A_-}:~~-X_0^2+X^2_1+...+X^2_{d+1}<0,~~X_0<0$, the region
lying inside the past light-cone ${\cal C}_-$ ($X_0<0$);

\medskip

\noindent 3.  ${\cal A_+}:~~-X_0^2+X^2_1+...+X^2_{d+1}<0,~~X_0>0$ , the region
lying inside the
future light-cone ${\cal C}_+$ ($X_0>0$).

\medskip

\noindent In each region Minkowski space can be foliated with
surfaces of constant curvature. In region ${\cal A}_-$ (or ${\cal
A}_+$) these are the surfaces described by the equation
\be
-X_0^2+X^2_1+...+X^2_{d+1}=-t^2
\lb{a}
\ee
for a constant $t$. The values of the time coordinate $t$ can be
arranged in such a way that region ${\cal A}_-$ is covered by
$-\infty < t<0$ and region ${\cal A}_+$ by $0<t<+\infty$. As is
well known, each of these surfaces with the metric induced from
(\ref{1}) is Euclidean anti-de Sitter space, which is a maximally
symmetric hyperbolic space. The metric on ${\cal A}$ foliated by
the surfaces (\ref{a}) is
\be
&&ds^2=-dt^2+t^2 ds^2_{H_{d+1}}~~, \nonumber \\
&&ds^2_{H_{d+1}}=dy^2+\sinh^2y d\omega^2_d
\lb{3}
\ee
where $ds^2_{H_{d+1}}$ is the standard metric on Euclidean anti-de
Sitter space; we adopt coordinates $(y,\theta)$ on $H_{d+1}$ such
that $y$ is the radial coordinate and $\{\theta\}$ are the angle
coordinates on the unit d-sphere with metric $d\omega^2_d$.

Similarly, outside the light-cone, region ${\cal D}$ can be sliced
with with de Sitter hypersurfaces,
\be
-X_0^2+X^2_1+...+X^2_{d+1}=r^2~~,
\lb{d}
\ee
which for any constant $r$ is a maximally symmetric Lorentzian
space with negative constant curvature. Using $r$ as a new
``radial'' coordinate we find that the Minkowski metric on this
foliation takes the form
\be
&&ds^2=dr^2+r^2ds^2_{dS_{d+1}}~~, \nonumber \\
&&ds^2_{dS_{d+1}}=-d\tau^2+\cosh^2\tau~d\omega^2_d
\lb{2}
\ee
where $ds^2_{dS_{d+1}}$ is the metric on $(d+1)$-dimensional
Lorentzian de Sitter space covered by the global coordinates
$(\tau,\theta)$, with $\tau$ playing the role of time and with
$\{\theta\}$ coordinates on the unit d-sphere.

Each of the metrics  (\ref{3}) and (\ref{2}) covers  only a part
of  Minkowski space. In order to cover the whole spacetime we need
metrics on three regions: ${\cal A}_-$, ${\cal A}_+$ and $\cal
D$\footnote{Strictly speaking, we also need to take the lightcone
itself into account. However, this is a subspace of measure zero,
and we will therefore ignore it, but it is indirectly present in
the form of the boundary conditions we will choose}. It is also
possible to introduce a closely related global coordinate system
$(x_+,x_-,\theta )$ which covers all of spacetime and in terms of
which the metric takes the well-known form \cite{HE}
\be
ds^2=-dx_+dx_-+{1\over 4}(x_+-x_-)^2d\omega^2_d~~.
\lb{metric}
\ee
In region $\cal A$ the relation between the coordinates
$(t,y,\theta )$ and the global coordinates is given by
\be
x_+x_-=t^2~,~~{x_-\over x_+}=e^{2y}~~.
\lb{coorA}
\ee
In region ${\cal A}_-$ both $x_+$ and $x_-$ are negative while in
region ${\cal A}_+$ they are both positive. In region $\cal D$ the
relation between the coordinates $(r,z,\theta )$ and the
coordinates $(x_+,x_-,\theta )$ reads
\be
x_+x_-=-r^2~,~~{x_+\over x_-}=-e^{2\tau}~~.
\lb{coorD}
\ee
One can see that in this region $x_+x_-<0$.
The light-cone ${\cal C}$ is defined by equation $x_+x_-=0$ so that its
component ${\cal C}_-$ corresponds to $x_+=0$ while the
equation for the future light-cone ${\cal C}_+$
is $x_-=0$.

It is interesting to determine the boundaries associated to the
slicing. First of all, the past light cone ${\cal C}_-$ has a
$d$-dimensional sphere $S^-_d$ as a boundary (lying at negative
infinite value of the time coordinate $X_0$). In terms of new
projective coordinates $\xi_1=X_1/X_0$, ...,
$\xi_{d+1}=X_{d+1}/X_0$ the equation of the sphere is
$$
\xi_1^2+...+\xi^2_{d+1}-1=0~~.
$$
Similarly, the boundary of the future cone ${\cal C}_+$ is a
sphere $S^+_d$. The hyperbolic surface (\ref{a}), as is well known
and much used in the context of the AdS/CFT correspondence, has a
d-dimensional sphere as a boundary. It is now important to note
that it is again the same sphere $S^-_d$. Indeed, in terms of the
coordinates $\{\xi_k\}$ equation (\ref{a}) becomes
$$
\xi_1^2+...+\xi^2_{d+1}-1=-t^2/X^2_0
$$
which (for any constant $t$) approaches the above written equation
for $S^-_d$ in the limit where $X_0 \rightarrow -\infty$.
Similarly, we can see that the boundary of a slice (\ref{a}) in
region ${\cal A}_+$ is the sphere $S^+_d$, the boundary of the
future cone ${\cal C}_+$. de Sitter space defined by equation
(\ref{d}) has two boundaries: a sphere lying in the infinite past
and a sphere lying in the infinite future. It is clear that for
any slice (\ref{d}) the boundaries are again respectively the
spheres $S^-_d$ and $S^+_d$. Thus the d-spheres $S^-_d$ and
$S^+_d$ are the only boundaries that appear in the slicing that we
are considering, and they therefore play an important role in the
remainder of the paper.
\begin{figure}
\centerline{\psfig{figure=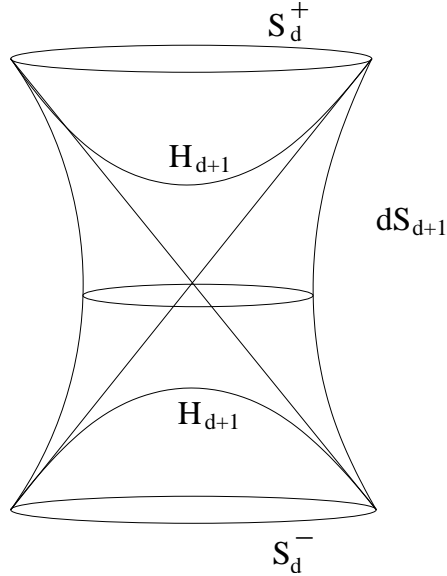}}
\caption{Inside light-cone (d+2)-dimensional Minkowski space-time  can be sliced with Euclidean anti-de
Sitter spaces $H_{d+1}$ while out-side the light-cone the appropriate
slices are Loretzian de Sitter spaces $dS_{d+1}$. The only boundaries of the slices
are  either  sphere $S^-_d$ or sphere $S^+_d$ lying at infinity of
the light-cone.}
\end{figure}
If we look at the initial value problem, we observe that the
hypersurfaces (\ref{a}) are not Cauchy surfaces. Initial data on
such a surface is not sufficient to determine the time-evolution
everywhere in Minkowski space. A complete set of initial data
should necessarily also include data on a de Sitter hypersurface
(\ref{d}). In this paper we shall however argue that all
sufficient information can be actually given on a lower
dimensional subspace:

\medskip

\noindent {\it  All information on (d+2)-dimensional Minkowski
space-time is holographically described by some data on the two
d-spheres $S_{d}^{-}$ and $S_{d}^{+}$}.

\medskip

\noindent In the following sections we shall give arguments in
support of this statement. Here we just note that this may not
seem to be too surprising. Indeed, we already know that all
information needed for anti-de Sitter space is contained in
certain data on its boundary. The same is true for de Sitter space
(at least classically) , only in this case the boundary has two
components, one at past and one at future infinity. In the
foliation of Minkowski space just described all (anti)-de Sitter
hypersurfaces have the same boundaries, $S_{d}^{-}$ and
$S_{d}^{+}$, and it is natural to expect that this is where all
needed information is stored.

Ending this section we  note that the isometry group O(d+1,1) of
Minkowski spacetime acts on the spheres $S^-_d$ (or $S^+_d$) as a
group of conformal symmetries. In section~6 we find explicitly the
form of the bulk diffeomorphisms generating conformal
transformations on the boundary of the light-cone. The conformal
structure associated with the  boundary of (anti-)de Sitter space
then appears to be a particular manifestation of the conformal
structure  associated with the infinity of the light-cone. The
relation between a Minkowski metric with Lorentzian signature, the
metric on Euclidean AdS and the conformal structure on the sphere
was earlier studied in \cite{FG}, a popular reference in the
literature on the AdS/CFT correspondence. The relevance of
Minkowski space has however not received much attention in
subsequent studies.

\section{Scalar Field in Minkowski space-time}
\setcounter{equation}0

Consider a scalar field with equation of motion
\be
\left( \nabla^2-M^2\right) \phi=0
\lb{5'}
\ee
on (d+2)-dimensional Minkowski spacetime  and use the slicing with
constant curvature spaces described in the previous section to see
whether the field can be equivalently described in terms of data
living on the slice. First we analyze the massless  case and take
the region ${\cal A_-}$ with metric (\ref{3}), $t$ being the
time-like coordinate taking negative values. The analysis for the
region ${\cal A_+}$ is similar. In terms of the metric (\ref{3})
the field equation (\ref{5'}) reads
\be
\Delta_{H_{d+1}}\phi -(d+1)t\partial_t \phi-t^2\partial^2_t\phi=0~~.
\lb{5''}
\ee
We can separate variables in this equation by taking $\phi$ of the
form $\phi\sim (-t)^{-\lambda (\mu )} \phi_\mu (y,\theta)$ where
$(y,\theta)$ is the set of coordinates on the Euclidean hyperbolic
space $H_{d+1}$ (which is the same as Euclidean anti-de Sitter
space). For a given $\mu$ the functions $\phi_\mu (x)$ satisfy the
massive field equation on $H_{d+1}$
\be
(\Delta_{H_{d+1}}-\mu^2)\phi_\mu (x)=0~~,
\lb{7}
\ee
and the parameter $\mu$ thus plays the role of mass. The original
$(d+2)$-dimensional field $\phi (t,x)$ reduces to a set of massive
fields on $(d+1)$-dimensional Euclidean anti-de Sitter space. The
relation between $\mu$ and the parameter $\lambda (\mu)$ can be
read from (\ref{5''}) and is given by\footnote{Not surprisingly,
the values of $\lambda_{\pm}$ are the same as the scaling
dimensions of the field theory operators corresponding to
$\phi_{\mu}$ in the AdS/CFT correspondence.}
\be
\lambda_\pm ={d\over 2} \pm \sqrt{\mu^2+{d^2\over 4}}~~.
\lb{lambda}
\ee
Thus, there are exactly two independent solutions,
$(-t)^{-\lambda_+}\phi^+(x) +(-t)^{-\lambda_-}\phi^-(x)$. When
$\mu^2+{d^2\over 4}=0$ the independent solutions are $t^{-{d\over
2}}$ and $t^{-{d\over 2}}\ln t$. Normally, we would choose one of
the solutions by imposing appropriate boundary conditions. One
possible condition is regularity at $t=0$ (i.e. near the past
light-cone ${\cal C}_-$) which selects only negative values of
$\lambda$. Note that at infinity (when $t$ goes to $-\infty$) we
pick up the most divergent solution. This is similar to the usual
story in anti-de Sitter space. Another possible condition is the
regularity of the action in a neighbourhood of $t=0$,
\be
W[\phi ]=\int_{t\rightarrow 0} t^{d+1} \phi (x,t)\partial_t \phi
(x,t)~~. \lb{action}
\ee
This gives the condition $Re \lambda <{d\over 2}$. In case $Re
\lambda={d\over 2}$ the expression (\ref{action}) does not contain
growing (as $t$ goes to zero) terms but there are oscillating
terms. These terms are allowed in quantum field theory since they
correspond to propagating waves, and they will also play an
important role in the remainder of this paper.

Note that the real part of $\lambda_+$ is always positive and thus
can not provide us with a regular solution at $t=0$. For
$\lambda_-$, depending on the value of $\mu^2$ in (\ref{7}), we
have three different cases:

1. $\mu^2>0$, then $\lambda_-$ is negative and the mode
$(-t)^{-\lambda_-}$ is regular at $t=0$.

2. $-{d^2\over 4}<\mu^2<0$, $\lambda_-$ is positive while still
real. There are no solutions regular at $t=0$ in this case.
However, $\lambda_-<{d\over 2}$ and hence the mode
$(-t)^{-\lambda_-}$ leads  to a regular action (at $t=0$).

3. $\mu^2<-{d^2\over 4}$, i.e. $\mu^2=-{d^2\over 4}-\alpha^2$,
then both $\lambda_+$ and $\lambda_-$ are complex:
$\lambda_\pm={d\over 2}\pm i\alpha$. The action has only
oscillating terms at $t=0$ in this case and both
$(-t)^{-\lambda_+}$ and $(-t)^{-\lambda_-}$ are allowed.

According to the AdS/CFT correspondence, to each field on anti-de
Sitter (with mass $\mu$) satisfying equation (\ref{7}) and the
Breitenlohner-Freedman bound $\mu^2>-d^2/4$ there exists a dual
conformal operator living on the boundary of anti-de Sitter
space (i.e. on the sphere $S^-_d$). The conformal dimension of the
operator is the largest root of the equation $h(h-d)=\mu^2$, for
positive $\mu^2$ it is
\be
h_+={d\over 2}+\sqrt{{d^2\over 4}+\mu^2}~~.
\lb{h}
\ee
Unitarity requires the conformal weight to be real  and that in turn implies
that $\mu^2$ has to satisfy the Breitenlohner-Freedman bound. We
thus see that this bound also has a clear meaning from the higher
dimensional Minkowski space prospective.

Notice that the propagating modes correspond to case 3 and are
thus described by complex conformal weights, $h_\pm={d\over 2}\pm
i \alpha$, with $\alpha$ an arbitrary real parameter. The standard
plane wave is described in terms of the metric (\ref{1}) as
$e^{ik_\mu X^\mu}$, where $k^2=0$. In the coordinates
$\{t,y,\theta\}$ it takes the form $e^{ip(y,\theta)t}$, where
$p(y,\theta)$ is some function of the coordinates  $y$ and
$\theta$. These modes can be decomposed in the basis formed by
functions $t^{-\lambda}$ where $\lambda$ is complex. This is seen
from the inverse Mellin type transform \cite{Shilov}
\be
e^{ipt}={1\over 2\pi i}\int_{c-i\infty}^{c+i\infty}d\lambda ~t^{-\lambda}
{i^\lambda \Gamma(\lambda)\over (p+i\epsilon)^\lambda} ~~,
\lb{Mellin}
\ee
where $\Gamma (\lambda )$ is the Gamma function. Thus, to describe
plane waves, or more generally any solution of the free field
equations which is normalizable in the sense of plane waves we
only need the modes with $\lambda={d\over 2}\pm i \alpha$.
Normally, such values of the conformal weights violate unitarity
in a dual CFT. However, it is important to keep in mind that we
are here discussing Euclidean AdS as a building block of a
Minkowski-signature theory, and we do not intend to analytically
continue the results to Lorentzian AdS. Therefore, it is not clear
whether the notion of unitarity that is normally inherited from
Lorentzian AdS has any relation to the notion of unitarity in
Minkowski space. Since Minkowski space physics carries a unitary
representation of the Lorentz group $SO(d+1,1)$, we would actually
expect that this is the group that is unitarily represented, and
for this group complex weights are not problematic, they
correspond to the principal series representation. All this is
very similar to a corresponding discussion for de Sitter space
given in \cite{Balasubramanian:2002zh}. We therefore view the
results in this paper as providing further motivation for the
claim in \cite{Balasubramanian:2002zh} that the appropriate notion
of unitarity in de Sitter space is one where $SO(d+1,1)$ is
unitarily implemented, and complex weights therefore do not
violate unitarity.

In the region ${\cal D}$ the story is essentially similar with the
coordinate $t$ replaced by the space-like coordinate $r$. Solving
equation (\ref{5'}) for the metric (\ref{d}) we decompose
solutions in terms of modes $r^{-\lambda}\phi_m (x)$, where
$\phi_m (x)$ is now a solution to the field equation with mass $m$
on de Sitter space
\be
(\Delta_{dS_{d+1}}-m^2)\phi_m=0~~.
\lb{10}
\ee
The relation between $m$ and $\lambda$ is given by
\be
\lambda_\pm={d\over 2}\pm \sqrt{{d^2\over 4}-m^2}~~.
\lb{l}
\ee
The analysis of modes, regularity conditions and conformal weights
are similar to the analysis in ${\cal A}_-$. To a large extent the
results can be obtained by analytical continuation of the
anti-de Sitter mass $m^2$ to the de Sitter mass $\mu^2=-m^2$. In
particular, the modes $r^{-\lambda_-}$ regular on the light cone
$(r=0)$ are the ones with $m^2<0$. In all versions of the dS/CFT
correspondence that have appeared in the literature the conformal
weight of the corresponding operator on the boundary of de Sitter
is
\be
h={d\over 2}+\sqrt{{d^2\over 4}-m^2}~~.
\lb{hd}
\ee
The modes with $0<m^2<{d^2\over 4}$ are not regular at the
light-cone but have finite action there. Finally, modes with
$m^2={d^2\over 4}+\alpha^2$ are oscillating at the light-cone.
Both values of the conformal weight, $h_+={d\over 2}+i\alpha$ and
$h_-={d\over 2}-i\alpha$, are allowed in this case. The latter
ones are the modes that are needed to describe plane-wave
normalizable solutions of the field equations.

The modes  described are analytic across the light-cone so that
modes with $m^2=\mu^2$ in region $\cal A$ become modes with
$m^2=-\mu^2$ in region $\cal D$. In particular, this means that
the limiting values $\{\phi_\mu [S^-]\}$ ($\{\phi_\mu [S^+]\}$) of
the wave function at $S^-_d$ ($S^+_d$) are the same regardless of
whether the light-cone is approached from region $\cal D$ or from
the region ${\cal A}_-$ (${\cal A}_+$). To see this explicitly
note that in region $\cal A_-$ close to the boundary at infinity
of the past light-cone the modes behave as
$(-t)^{-\lambda_-}e^{-\lambda_- y}=(-x_-)^{-\lambda_-}$, where
$x_-$ (see (\ref{coorA})) is a coordinate in the coordinate system
(\ref{metric}) covering all of spacetime. It approaches $-\infty$
as we approach the boundary at infinity of the past light-cone. On
the other hand, in the region $\cal D$ the modes (with
$m^2=-\mu^2$, i.e. with the same $\lambda_-$) behave in the same
region as $r^{-\lambda_-}e^{\lambda \tau}=(-x_-)^{-\lambda_-}$,
where the equality follows from (\ref{coorD}). The past light-cone
is the surface where $x_+=0$. We see that the modes are analytic
across this surface and describe a wave configuration propagating
along the past light-cone. In a similar fashion one finds that
the modes are also analytic across the future light-cone.

Let us now turn on the mass in  equation (\ref{5'}). The
decomposition of the bulk field over a set of massive fields
living on the slice is valid also in this case. However the radial
functions in the decomposition (we take region $\cal D$ for
concreteness)
\be
\phi (r,x)=\int dm {1\over r^{d/2}}K_{\nu (m)}(Mr)~\phi_m (x)
\lb{mass}
\ee
are now expressed in terms of the MacDonald
function\footnote{Another possible Bessel function (Infeld
function $I_\nu (Mr)$) is exponentially growing at infinity and
thus omitted.} $K_\nu (Mr)$, where $\nu=\sqrt{{d^2\over 4}-m^2}$.
The integration over $m$ in (\ref{mass}) should be understood as
symbolic, the range of values and measure on that space may depend
on the class of functions we want to consider. Each massive field
on the slice has a holographic description in terms of conformal
operators on the boundary of the slice. The relation between the
conformal weight and the mass $m$ is given by (\ref{hd}). In this
sense there is no difference with the massless bulk field
considered above. A plane wave solution of the massive field
equation in the bulk can also be decomposed in terms of the radial
functions (\ref{mass}). The relevant integral transform is known
in the mathematical literature as the Kantorovich-Lebedev
transform. The analog of equation (\ref{Mellin}) is
\be
e^{ipr}=\int_0^{+\infty}d\alpha ~r^{-d/2} K_{i\alpha}(Mr) f(p,\alpha)
\lb{Kant-Leb}
\ee
where the function $f(p,\alpha )$ is expressed in terms of a
hypergeometric function as follows
\be
f(p,\alpha)={2\alpha\over \pi^{3\over 2}}\sinh\alpha {1\over (2M)^{d\over 2}}
{\Gamma ({d\over 2}+i\alpha)\Gamma ({d\over 2}-i\alpha)\over \Gamma ({d+1\over 2})}
F({d\over 2}+i\alpha, {d\over 2}-i\alpha,{d+1\over 2}, ({1\over 2}+{ip\over 2M}))~~.
\lb{fp}
\ee
Interestingly, it can be brought to a compact form
\be
f(p,\alpha)={2(2\pi )^{d/2}\over \Gamma (i\alpha )\Gamma (-i\alpha )}M^{-d/2} ~
D_{{d\over 2}+i\alpha}(i{p\over M})~~
\lb{KL1}
\ee
in terms of a de Sitter invariant Green's function $D_\lambda (z)$
to be defined in section 5 ($z$ in that case would be related to
the geodesic distance on a de Sitter slice). The appearance of
this function in the Kantorovich-Lebedev transform
(\ref{Kant-Leb}) is rather mysterious. Since massive fields in the
bulk do not lead to any immediate new conceptual issues but are
mainly technically more complicated, we will restrict our
attention to massless fields in the remainder of this paper.

Summarizing our progress so far, we have found that the scalar
(massive or massless) field propagating in Minkowski space is
equivalently described in terms of an infinite set of massive
fields living (depending on the region of Minkowski space) on
anti-de Sitter or de Sitter space. The original scalar field can
thus be decomposed in terms of massive fields on (anti-) de
Sitter. This is a sort of spectral decomposition since it involves
an integration over the mass $\mu$ (or $m$), which therefore
plays the role of the spectral parameter. The explicit form of
this decomposition will be studied in section~5.

Several of the equations and decompositions given in this and
later sections have also been obtained in the papers
\cite{Bertola,Bertola2}, but we are not aware of any attempt to
use de Sitter and AdS slices to try to find a holographic dual
description of Minkowski space. Similarly, one can consider
massless higher spin fields, but the generalization is rather
straightforward and will not be discussed in this paper.

As we know from the AdS/CFT correspondence any massive field on
Euclidean anti-de Sitter space is completely described by fixing
Dirichlet data on the boundary of the space, which in the present
case is the sphere $S^-_d$ (or $S^+_d$ for the region ${\cal
A}_+$). One has to be a bit careful in applying the standard
AdS/CFT formalism, because in our case we have complex conformal
weights and there is no longer a separation of fields in
normalizable and non-normalizable modes. The analog statement for
de Sitter space is that the initial-time data can be fixed on
either the past or future infinity surface which is again
respectively $S^-_d$ or $S^+_d$. This suggests that the Cauchy
problem in Minkowski space becomes the problem of reconstructing
the values of a field everywhere in terms of known data
$\{\phi_\mu [S^-]\}$ given on the past infinity ($S^-_d$) of the
light cone. The explicit form of this reconstruction will be given
in subsection~5.2. Having in mind the Quantum Mechanical picture
we would expect that the quantum in-state should be associated
with data on $S^-_d$ while the out-state should be defined in
terms of the data $\{\phi_\mu [S^+]\}$ on the future infinity
($S^+_d$) of the light cone. In the next sections we will
elaborate more on this picture.

\section{Light-cone conformal correlation functions}
\setcounter{equation}0
In this section we show that in a manner similar to the AdS/CFT correspondence
\cite{Wit} there naturally appear conformal field theory correlation functions on
 the boundary of the light-cone in Minkowski spacetime. Actually, the discussion
in this section should be taken with a grain of salt and be viewed as providing
some inspiration for the results in the remainder of the paper.

Consider the action of the scalar field on ${\cal A_-}$. If the
field is a solution of the equations of motion the action reduces
to the boundary term
\be
W=\int_{t\rightarrow -\infty}  (-t)^{d+1}\sqrt{\gamma} \phi (x, t)
\partial_t \phi (x,t)~~, \lb{12'}
\ee
where $x\equiv (y,\theta )$ are coordinates on the Anti-de Sitter
slice, $H_{d+1}$. We decompose fields in terms of modes as we did
in the previous section,
\be
\phi(x,\tau )=\sum_{(h_+,h_-)}(-t)^{-h_+}\phi^{(+)}(x)+
(-t)^{-h_-}\phi^{(-)}(x)~~,
\lb{dec}
\ee
where we sum (or, rather, take an integral if $h_\pm$ are
continuous) over all possible pairs $(h_+,h_-)$. In terms of the
massive fields (with mass $\mu$) on anti-de Sitter space
$H_{d+1}$, $h_\pm$ are the roots of the equation $h(h-d)=\mu^2$
($h_++h_-=d$, $\mu^2=-h_+h_-$). In this section we do not specify
the sign of $\mu^2$. In the decomposition (\ref{dec}) a priori all possible
(both real and complex) conformal weights should be taken into
account, but, as we discussed in the previous section, plane wave normalizability
restricts the conformal weights to $h_{\pm}=\frac{d}{2} \pm i\alpha$.
Next, we substitute (\ref{dec}) into (\ref{12'}) and take
the limit $t\rightarrow -\infty$. In general the resulting expression has
infinite,  oscillating and finite terms
$$
W=\sum_{h+h'=d}\int_{H_{d+1}}h'\ \phi_h(x)\phi_{h'}(x)+
\sum_{{\rm Re}(d-h-h')>0}h'\ (-t)^{(d-h-h')}\int_{H_{d+1}}\phi_h(x)\phi_{h'}(x)~~
$$
where the infinite and oscillating terms are collected in the second sum.
We are interested in
the finite one,
\be
W_{fin}=\sum_{(h_+,h_-)} d \int_{H_{d+1}} \phi^{(+)}(x)\phi^{(-)}(x)~~,
\lb{13'}
\ee
where  the
summation (integration) runs over all possible pairs $(h_+,h_-)$.

Before proceeding, we point out two problems with the analysis in
this section. First, we have to drop some infinite and oscillating 
terms in order to get
(\ref{13'}). In the spirit of the usual AdS/CFT correspondence, such
a procedure would be viewed as the holographic version of renormalization,
but here such an interpretation is problematic, because the subleading
terms in the $t$ expansion are non-local functionals of the leading term,
and therefore the infinities are non-local in nature \cite{deHaro:2000wj}.
However, what we really want to do is holography in each of the AdS slices
and keep the $t$-dependence exactly, and then this issue will not arise.
Besides this, there are problems evaluating on-shell actions for
scalar fields on de Sitter slices, and probably also for fields on
Euclidean anti-de Sitter space with complex conformal weights, as discussed
in \cite{Balasubramanian:2002zh}. Both problems will not arise in the
next section where we use an S-matrix type approach to the Minkowski/CFT duality,
and as we said above, this section should mainly be viewed as providing
some inspiration.

Continuing with the analysis, we recall that
each field $\phi^{(+)}(x)$ and $\phi^{(-)}(x)$ is a solution of the
scalar field equation
$$
(\Delta -\mu^2)\phi^{(\pm )}(x)=0
$$
with mass $\mu^2=-h_+h_-$ on $H_{d+1}$. The coordinates
$x=(\theta,y)$ are such that $y=\infty$ corresponds to the
boundary of $H_{d+1}$  and $(\theta )$ are the angle coordinates
on the asymptotic sphere $S^-_d$.

Near the boundary of $H_{d+1}$ there exists a decomposition
$$
\phi^{(+)}(x)=e^{-h_-y}\phi^{(+)}_-(\theta )+e^{-h_+y}\phi^{(+)}_+(\theta)
$$
$$
\phi^{(-)}(x)=e^{-h_-y}\phi^{(-)}_-(\theta )+e^{-h_+y}\phi^{(-)}_+(\theta )
$$
familiar from the AdS/CFT duality \cite{Wit}. If $h_+$ and $h_-$
are real the term $e^{-h_-y}$  dominates and the functions
$\phi^{(+)}_- (\theta)$, $\phi^{(-)}_- (\theta)$ are the usual
choices for the Dirichlet boundary condition at the boundary of
AdS space $H_{d+1}$. If the conformal weights are complex any
function $\phi^{(\pm)}_- (\theta)$ or $\phi^{(\pm)}_+ (\theta)$
can be used as a boundary condition. The solutions of the massive
scalar field equation on $H_{d+1}$ can then chosen to be
\be
&&\phi^{(+)}(\theta,y)=\int_{S_d} d\mu(\theta')~G_- (y,\theta,\theta')~\phi^{(+)}_+ (\theta')~~,
\nonumber \\
&&\phi^{(-)}(\theta,y)=\int_{S_d} d\mu(\theta')~G_+ (y,\theta,\theta')~\phi^{(-)}_- (\theta')~~,
\lb{14'}
\end{eqnarray}
where $d\mu(\theta)$ is the standard measure on a unit
radius d-sphere, and we will from now on use the notation $\gamma
(\theta,\theta')$ for the geodesic distance on a d-sphere between
points with coordinates $\theta $ and $\theta'$.
We will ignore the fact that for complex weights the fields in (\ref{14'})
are generically complex, and one should really add the complex conjugate fields
in the right hand side. Again, the results here should be taken with a grain
of salt.
The kernels
\be\lb{14''}
&&G_h (y,\theta,\theta' )=
{g(h)\over (\cosh y-\sinh y \cos \gamma (\theta,\theta'))^{h}}~~,
\nonumber \\
&&g(h)={(4\pi )^{-{d\over 2}}}~{\Gamma(h) \over \Gamma (h-{d\over 2})}~~,
\end{eqnarray}
can be interpreted as bulk-boundary propagators on an anti-de
Sitter slice. For large $y$ they behave as follows (see Appendix \ref{kernels}
for details)
\be
&&\lim_{y\to\infty}G_+(y,\theta ,\theta')=(e^{-h_-y}\delta^{(d)}(\theta,\theta ')+e^{-h_+y}K_+(\theta,\theta'))(1+{\cal O}(e^{-2y})) \nonumber \\
&&\lim_{y\to\infty}G_-(y,\theta ,\theta')=(e^{-h_+y}\delta^{(d)}(\theta,\theta ')+e^{-h_-y}K_-(\theta,\theta'))(1+{\cal O}(e^{-2y}))~~,
\lb{G}
\ee
where $\delta^{(d)}(\theta,\theta ')$ is the delta-function on the d-sphere.
The kernels $K_-(\theta,\theta')$ and $K_+(\theta,\theta')$
are defined as
\be
K_\pm (\theta,\theta')
={2^{h_\pm}~g(h_\pm)\over (1-\cos\gamma (\theta,\theta'))^{h_\pm}}
\lb{K}
\ee
and are inverse to each other
\be
\int d\mu (\theta) K_-(\theta,\theta')K_+(\theta,\theta'')=\delta^{(d)}(\theta',\theta'')~~.
\ee
Thus, from (\ref{14'}) we have that
\be
&&\phi^{(\pm)}_- (\theta)=\int d\mu(\theta ')~K_-(\theta,\theta')
~\phi^{(\pm)}_+ (\theta') \nonumber \\
&&\phi^{(\pm)}_+ (\theta)=\int d\mu(\theta ')~K_+(\theta,\theta')~
\phi^{(\pm)}_- (\theta') ~~.
\lb{Kphi}
\ee
The idea is to substitute (\ref{dec})-(\ref{14'}) into the action (\ref{12'}),
first take the integration over $y$  and then take the limit  of infinite $t$.
Note, that  additionally to the divergences at large $t$
there could  be extra
divergences when the integration over $y$ is being taken.

Another important point is to observe that
\be
\int_{S_{d}}d\mu(\theta') \int_0^\infty dy ~\sinh^d(y) \
G_{h}(y,\theta',\theta)\ G_{h'}(y,\theta',\theta'')
\sim~\delta^{(d)}(\theta,\theta'')\delta(d-h-h')~~, \lb{15'}
\ee
where we assumed that $h$ and $h'$ take values on the complex
line $d/2+i\alpha$, and possibly omitted some regular terms in the
right hand side. We suspect these regular terms are not there, but
have not been able to prove this. 
The delta-function  in
(\ref{15'}) can be recognized in the asymptotics (\ref{G}) of the
kernels $G_h (y,\theta, \theta')$ (note, that 
we also omitted the terms proportional to $\delta(h-h')$ which 
contribute to the oscillating part of the action). With the help of this formula
we find that the finite part of the action, after the integrations
over $t$ and $y$ have been performed, is given by
\begin{eqnarray}
&&W_{fin}\propto \sum_{(h_++h_-=d)} \int_{S^{-}} \phi^{(-)}_{-}
(\theta)\phi^{(+)}_{+} (\theta)
\nonumber \\
&&=\sum_{(h_+ + h_-=d)} \int_{S^-}\int_{S^-} d\mu (\theta)~
\phi^{(-)}_{-} (\theta)~ K_{+}(\theta,\theta')~ \phi^{(+)}_{-}
(\theta')~d\mu (\theta')~~. \lb{16'}
\end{eqnarray}
For complex conformal weights we have
$[\phi^{(-)}_{-}]^*=\phi^{(+)}_{+}$ and $[\phi^{(-)}_{+}]^*=\phi^{(+)}_{-}$
so that various complex conjugates will appear in (\ref{16'}).
If we were to follow the usual AdS/CFT story, we would interpret
the
functions $\phi^{(-)}_-(\theta)$ and $\phi^{(+)}_-(\theta)$ as
sources coupled to dual conformal operators
${\cal O}^{(-)}_+(\theta)$ and ${\cal O}^{(+)}_+(\theta)$ respectively
(both with conformal weight $h_+$).
Equation (\ref{16'}) would then indicate that there is a  non-trivial
correlation function between the  operators
$O^{(+)}_+ (\theta)$ and $O^{(-)}_+(\theta)$ and that it takes
the form of a conformal two-point function on the d-sphere.
Again, this is all suggestive but quite imprecise.

A similar analysis can be done  in regions ${\cal A}_+$ and
${\cal D}$. By finally considering the action on all of Minkowski space
we arrive at a functional
$
W_{fin}[\phi_{h}[S^-],\phi_{h}[S^+]]
$
of
data
$\phi_{h}[S^-]$ and $\phi_{h}[S^+]$
on the two d-spheres $S^-$ and $S^+$
respectively.
The variation with respect to $\phi_{h}[S^-]$ and then with respect to
$\phi_{h}[S^+]$
is naturally interpreted as a correlation function between operators living on
$S^+$ and $S^-$. The interesting question arises whether
those correlation functions have anything to do with the S-matrix
in Minkowski space (once we include interactions).
We will return to this and other questions in the next section,
where we will set up a more precise version of the Minkowski/CFT
correspondence.

\section{Green's functions and  S-matrix in \\
 Minkowski spacetime}
\setcounter{equation}0

\subsection{Propagators }

The manifestation of the conformal structure of correlation
functions associated with the asymptotic boundaries of the
light-cone which we studied in the previous section was
suggestive.  The emergence of conformal structures becomes however
clearer and more precise if we study the asymptotic behavior of
the propagator in Minkowski spacetime.

For simplicity we start with the free field and consider the Hadamard  type
propagator
defined as
$$
D(X,X')=<0|\{\phi (X),\phi (X') \}|0>~~.
$$
In (d+2)-dimensional Minkowski space it takes the form
\be
D(X,X')={\Omega_d\over (s^2)^{d/2}}~~, ~~s^2=-(X_0-X_0')^2+ (\bf{X}-\bf{X}')^2~~,
\lb{p0}
\ee
where $s^2$ is the space-time interval between two events and $\Omega_d=
{\Gamma({d\over 2})\over 4\pi^{d/2+1}}$.

Let both points lie in the region $\cal D$ foliated by de Sitter
slices. The foliation is given by coordinates $(r,\tau,\theta )$
that are related to the coordinates of Minkowski space as follows
\be
X_0=r\sinh\tau~~,~~{\bf{X}} = r\cosh\tau ~{\bf{n}}(\theta)~~,
\lb{X}
\ee
where ${\bf{n}}(\theta)$ is d-dimensional unit vector parametrized
by angle coordinates $\{\theta \}$.
The distance between two points with coordinates $(r,\tau,\theta)$
and $(r',\tau',\theta' )$ is
\be
s^2=r^2+r'^2-2rr'\cos \sigma~~,
\lb{p1}
\ee
where $\sigma$ is the geodesic distance on de Sitter space between
the points $(\tau,\theta )$ and $(\tau',\theta')$
\be
\cos\sigma=-
(\sinh\tau \sinh\tau'-\cosh\tau\cosh\tau'\cos\gamma(\theta, \theta'))~~,
\lb{p2}
\ee
where $\gamma(\theta, \theta')$ is the angle between the unit
vectors ${\bf n}(\theta)$ and ${\bf n}(\theta')$, which at the
same time is the geodesic distance on the d-sphere. Note that
$\sigma^2>0$ for space-like intervals on de Sitter space while
$\sigma^2<0$ for time-like intervals. In the latter case
$\cos\sigma$ in (\ref{p1}) and (\ref{p2}) becomes $\cosh |\sigma
|$.

Substituting (\ref{p1}) into (\ref{p0}) and expanding in powers of $r/r'$
we recognize a  generating function for Gegenbauer polynomials
$$
(1-2(r/r')\cos\sigma  +({r/r'})^2)^{-\nu}=
\sum_{n=0}^\infty C^\nu_n(\cos\sigma )(r/r')^n~~.
$$
The infinite sum can be represented as a single contour integral
so that we have the following  useful representation for the
Hadamard function
\be
D(X,X')=-{i\over 2}\Omega_d \int_{c-i\infty}^{c+i\infty} r^{-\lambda}r'^{\lambda-d}
C^{d/2}_{-\lambda} (-\cos\sigma){d\lambda\over \sin(\lambda\pi)}~~,
\lb{p3}
\ee
where the constant $c$ should satisfy the condition $0<c<d$. We
can choose $c=d/2$ so that the integral is over
$\lambda=d/2+i\alpha$ where $\alpha$ changes from minus to plus
infinity. These values of $\lambda$ are natural because they
appear in the decomposition of plane waves as we saw in section~3.

The propagator is now represented in terms of Gegenbauer functions
which for arbitrary $\lambda$ are defined in terms of the
hypergeometric function as
$$
C^{d/2}_{-\lambda} (-z)={\Gamma (d-\lambda )\over \Gamma (1-\lambda)\Gamma(d)}
F(d-\lambda,\lambda;{d+1\over 2};{1\over 2}(1+z))~~.
$$
When two points on a slice approach each other the geodesic
distance $\sigma$ goes to zero and the argument in Gegenbauer
function approaches -1. Therefore, the argument in the
hypergeometric function goes  to $+1$. This is the point where the
hypergeometric function  has a singularity,
$F(\alpha,\beta;\gamma;z)\sim
(1-z)^{\gamma-\alpha-\beta}~~,~z\rightarrow 1$. For the
hypergeometric function in the definition of Gegenbauer function
we have $\gamma-\alpha-\beta=-(d-1)/2$. Thus, when $\sigma$
approaches zero Gegenbauer function in (\ref{p3}) has singularity
$$
C^{d/2}_{-\lambda}(-\cos\sigma)\sim {1\over \sigma^{d-1}}
$$
which depends only on the dimension $d$ but not on  $\lambda$.
This is exactly the expected behavior of  Green's function.
Indeed, the Gegenbauer function $C^{d/2}_\lambda (-\cos\sigma)$ is
related to the   de Sitter invariant Green's function defined as
$$
D_\lambda (z)={\Gamma (\lambda)\Gamma(d-\lambda)\over (4\pi )^{d+1\over 2}
\Gamma({d+1\over 2})}F(d-\lambda,\lambda, {d+1\over 2},{1\over 2}(1+z))~~,
$$
where $z=\cos\sigma$, of the field operator $(\Delta-\mu^2)$,
$\mu^2=\lambda(\lambda-d)$. The exact relation is
\be
C^{d/2}_{-\lambda}(-z)={4\pi^{d/2}\over \Gamma({d\over 2})}\sin (\pi\lambda) ~
D_\lambda (z)~~.
\lb{CD}
\ee
Obviously, $D_\lambda (z)=D_{d-\lambda}(z)$. Equation (\ref{p3})
then takes the especially simple form
\be
D(X,X')={1\over 2\pi i}\int_{c-i\infty}^{c+i\infty}
d\lambda~ r^{-\lambda}r'^{\lambda-d}
D_\lambda (\cos\sigma)
\lb{DD}
\ee
which shows how the Green's function in $(d+2)$-dimensional
Minkowski spacetime is decomposed in terms of the infinite set of
Green's functions on the $(d+1)$-dimensional de Sitter slice
parametrized by $\lambda$.

In region $\cal A$, inside the light-cone, we choose the
coordinate system $(t,y,\theta )$ as follows
\be
X_0=t\cosh y~~,~~ {\bf{X}} = t\sinh y ~{\bf{n}}(\theta)~~.
\lb{X1}
\ee
The Minkowski distance in this region  takes the form
\be
s^2=-(t^2 +t'^2-2tt'\cosh\sigma )~~,
\lb{p4}
\ee
where $\sigma$ is the geodesic distance on Euclidean anti-de
Sitter space defined as
\be
\cosh\sigma =\cosh y \cosh y'-\sinh y \sinh y' \cos \gamma (\theta ,\theta')
\lb{p5}
\ee
and as before $\gamma (\theta ,\theta')$ is the geodesic distance
on the unit d-sphere. The propagator in this region has a
decomposition similar to (\ref{DD}),
\be
D(X,X')={(-i)^{d} \over 2\pi i}
\int_{c-i\infty}^{c+i\infty}d\lambda~ t^{-\lambda} t'^{\lambda-d}
D_\lambda (\cosh\sigma) ,\lb{p6}
\ee
in terms of an infinite set of Green's functions $D_\lambda
(\cosh\sigma)$ on the anti-de Sitter slices.

In Quantum Field Theory a special role is played by the Feynman
propagator which in Minkowski space is given by
\be
D_{\tt F}(X,X')=\Omega_d~{i\over (s^2+i\epsilon )^{d/2}}
\lb{p10}
\ee
with the appropriate $i\epsilon$-prescription showing that the
propagator is analytic in the upper half-plane as a function of
the invariant distance $s^2$. This translates directly into a
corresponding prescription for the propagators on the de Sitter
slices once we use the decomposition (\ref{p3}). We find for the
Feynman-like propagator $D_\lambda^{\tt F}$ as function of the
de-Sitter invariance distance $\sigma$ that
$$
D_\lambda^{\tt F}(\cos\sigma)=i D_\lambda (\cos\sigma-i\epsilon )~~.
$$
This gives the correct Feynman prescription for going around the
$\sigma=0$ singularity in the propagator on de Sitter spacetime.

It is now our goal to see how the Feynman propagator in Minkowski
space behaves when the two points approach the boundaries at
infinity of the light-cone. Suppose that one of the points
approaches the future infinity of the light-cone, i.e. the sphere
$S^+$, from the side of the region $\cal D$. In this case we have
$\tau'\rightarrow +\infty$ and the de Sitter geodesic distance
behaves in this limit as
$$
z=\cos\sigma\sim{1\over 2}e^{\tau'}(-\sinh\tau+\cosh\tau\cos\gamma
(\theta,\theta'))~~.
$$
Using the known properties (available in \cite{GR}) of the
hypergeometric function when its argument approaches infinity we
find that as $\tau'\rightarrow +\infty$
$$
D_\lambda (z)\sim{\Gamma (\lambda)\Gamma (d-2\lambda)\over
(4\pi)^{d+1\over 2} \Gamma({d+1\over 2}-\lambda)}{2^\lambda\over
(-z)^\lambda}+(\lambda\leftrightarrow d-\lambda )~~.
$$
This expression contains factors $(\sinh\tau -\cosh\tau \cos\gamma
(\theta,\theta')+i\epsilon)^{-\lambda}$. When now the second point
is taken either to the future infinity ($\tau\rightarrow +\infty$)
or to the past infinity ($\tau\rightarrow- \infty)$ of the
light-cone these factors produce non-trivial distributions on the
d-sphere (see Apendix A for details)
\be
&&\lim_{\tau\to+\infty}
(\sinh\tau -\cosh\tau \cos\gamma (\theta,\theta')+i\epsilon)^{-\lambda}
\nonumber \\
&&=(4\pi )^{d/2} ~i^{(d-2\lambda)}~{\Gamma(\lambda-{d\over 2})\over
\Gamma(\lambda)} e^{-(d-\lambda)\tau}\delta^{(d)}(\theta,\theta')+
{2^\lambda e^{-\lambda\tau}\over (1-\cos\gamma(\theta,\theta'))^\lambda}
\lb{Glambda}
\ee
and
\be
&&\lim_{\tau\to-\infty}
(\sinh\tau -\cosh\tau \cos\gamma (\theta,\theta')+i\epsilon)^{-\lambda}
\nonumber \\
&&=(4\pi )^{d/2} ~(-i)^{d}~{\Gamma(\lambda-{d\over 2})\over
\Gamma(\lambda)} e^{(d-\lambda)\tau}\delta^{(d)}(\theta,\pi-\theta')+
{2^\lambda (-i)^{2\lambda}e^{\lambda\tau}\over (1+\cos\gamma(\theta,\theta'))^\lambda}~~,
\lb{Glambda1}
\ee
where $ \pi -\theta $ stands for  the spherical coordinates of the
anti-podal point of the d-sphere, in other words ${\bf
n}(\theta)=-{\bf n}(\pi-\theta)$. The expressions in
(\ref{Glambda1}) are valid up to terms of order $e^{-2\tau}$.
Using (\ref{Glambda1}) we find the form of the Feynman propagator
when both points approach the future infinity of the light-cone
($\tau,\tau'\rightarrow +\infty$)
\be
&&\lim_{\tau,\tau'\to +\infty}D_{\tt F}(X,X')
=\int {d\lambda\over 2\pi i} r^{-\lambda}
r'^{-(d-\lambda)}\nonumber \\
&&\left( A(\lambda )
{e^{-\lambda (\tau'+\tau)}\over (1-\cos \gamma )^\lambda}
+B(\lambda )\delta^{(d)} (\theta,\theta')e^{-(d-\lambda)\tau} e^{-\lambda\tau'}
+(\lambda\leftrightarrow d-\lambda)\right)
\lb{p8}
\ee
where the integration over $\lambda$ goes from $d/2-i\infty$ to $d/2+i\infty$.
A similar expression appears when $\tau,\tau'\rightarrow -\infty$.

Notice that the first term in (\ref{p8}) looks like a conformal
correlation function on one of the boundaries ($S^+_d$) of the
light-cone, similar to the expression we found in the previous
section. From a holographic point of view, it is a correlation
function between operators associated with modes
$r^{-\lambda}e^{-\lambda \tau}$ and $r^{-(d-\lambda)}e^{-\lambda
\tau}$. On the other hand, we find that the correlation functions
of the operators associated with the modes
$r^{-\lambda}e^{-(d-\lambda)\tau}$ and
$r^{-(d-\lambda)}e^{-\lambda\tau}$ produce contact terms
proportional to a delta-function on the d-sphere.

When one of the points approaches the future infinity of the light-cone
and another approaches the past infinity of the  light-cone we find that
\be
&&\lim_{\tiny
\begin{array}{c}  \tau' \to +\infty \\ \tau \to -\infty \end{array}}
D_{\tt F}(X,X')=\int {d\lambda \over 2\pi i} r^{-\lambda}
r'^{-(d-\lambda )}\nonumber \\
&&\left(D(\lambda )
{e^{\lambda (\tau-\tau')}\over (1+\cos \gamma )^\lambda}+
F(\lambda )\delta^{(d)} (\theta,\pi-\theta')e^{(d-\lambda)\tau} e^{-\lambda\tau'}
+(\lambda\leftrightarrow d-\lambda)\right)~~
\lb{p9}
\ee
This also takes the form of correlation functions between
conformal operators living on $S^-_d$ and $S^+_d$: the correlation
function diverges for antipodal points on d-sphere. This is
similar to the behavior for two point functions in de Sitter space
as discussed in \cite{Str,st2,Spradlin:2001nb}. The functions
$A(\lambda)$, $B(\lambda)$, $D(\lambda)$ and $F(\lambda)$ are
expressed in terms of the Gamma function as follows
\be
&&A(\lambda)={2^\lambda i \over 4\pi^{{d\over 2}+1}}
\Gamma (\lambda )\Gamma({d\over 2}-\lambda)~,~~B(\lambda)={2^d i^{(d-2\lambda+1)}\over 4\pi}\Gamma({d\over 2}-\lambda)
\Gamma(\lambda-{d\over 2}) ~~, \nonumber \\
&&D(\lambda)={2^\lambda (-i)^{(2\lambda-1)}\over 4\pi^{{d\over 2}+1}}\Gamma (\lambda )\Gamma({d\over 2}-\lambda)~,~~
F(\lambda)={2^{d}(-i)^{(d-1)}\over 4\pi}\Gamma({d\over 2}-\lambda)
\Gamma(\lambda-{d\over 2})~~.
\lb{ABDF}
\ee

We can repeat this analysis in the case when  either of the  infinities
of the light-cone is
approached from the side of the region $\cal A$.
It is useful to note that the coordinate systems (\ref{X})
and (\ref{X1}) are related by the analytic transformation
\be
&&{\cal D}\rightarrow {\cal A}_+:r\rightarrow i t~~,~~\tau\rightarrow y-i{\pi\over 2}~~,~~\theta\rightarrow \theta \nonumber \\
&&{\cal D}\rightarrow{\cal A}_-:r\rightarrow -i t~~,
~~\tau\rightarrow -y+i{\pi\over 2}~~,~~\theta\rightarrow \pi-\theta~~.
\lb{p7}
\ee
Using the analytic continuation (\ref{p7}) between regions $\cal
A$ and $\cal D$ one can demonstrate that the structure of the
conformal correlation functions (\ref{p8}) and (\ref{p9}) is the
same no matter from which side each point approaches the boundary
of the light-cone. Thus, the structure of the conformal
correlators emerging in (\ref{p8}) and (\ref{p9}) is an intrinsic
property of the boundaries ($S^-_d$ and $S^+_d$) themselves.

Incidentally, in de Sitter space there is a one-parameter family
of vacua usually called $\alpha$-vacua, and one may wonder which
vacuum state is natural once de Sitter space is embedded in
Minkowski space. According to \cite{Bertola2}, the natural de
Sitter vacuum to use in this context is the Bunch-Davies vacuum or
the Euclidean vacuum. We have not verified whether the form of
(\ref{p9}) is consistent with the two-point functions in the
Euclidean vacuum as derived in \cite{st2,Spradlin:2001nb}, or
whether it corresponds to a different value of $\alpha$, but
clearly it would be interesting to investigate this.

\subsection{In- and out-fields}

The standard decomposition of asymptotic fields  in terms of plane waves
is
\be
\phi(X_0,{\bf X})=\int d{\bf k}\left(a({\bf{k}}) f_k(X_0,{\bf{X}})
+a^{\dagger}({\bf{k}})f^*_k(X_0,{\bf{X}})\right)~~,
\lb{p11}
\ee
where the functions
\be
f_k(X_0,{\bf{X}})={1\over [(2\pi)^{d+1}2k]^{1/2}}e^{-i(kX_0-\bf{k}\bf{X})}
\lb{fk}
\ee
form a basis of (positive frequency) asymptotic fields. The
(d+1)-vector ${\bf k}$ is parametrized by its length
$k=\sqrt{{\bf{k}}{\bf{k}}}$ and angles $\{\theta_{k}\}$ which
determine the direction of the vector in momentum space. Hence we
have ${{\bf k}}=k~ {\bf{n}} (\theta_k)$, where
${\bf{n}}(\theta_k)$ is unit vector providing the direction. The
creation and annihilation operators satisfy the usual algebra
\be
[a({\bf{k}}),a^{\dagger}({\bf{k'}})]=\delta^{(d+1)}({\bf{k}}, {\bf{k'}})={1\over k^d}\delta(k-k')~
\delta^{(d)}(\theta_k,\theta_{k'})~~,
\lb{p12}
\ee
where $\delta^{(d)}(\theta_k,\theta_{k'})$ is the delta-function
on the d-sphere with respect to the standard measure $d\mu
(\theta_k)$.

In the region $\cal A$ the coordinates $(X_0, \bf{X})$ are given by (\ref{X1})
so that the argument in the  exponent in (\ref{p12}) can be written in the form
$$
(kX_0-{\bf{k}}{\bf{X}})=kt~(\cosh y-\sinh y\cos\gamma(\theta,\theta_k))~~.
$$
Using this expression and formula (\ref{Mellin}) we
obtain a representation of the plane waves
\be
e^{\pm i(kX_0-{\bf{k}}{\bf{X}})}={1\over 2\pi i}\int_{c-i\infty}^{c+i\infty}
d \lambda {(\pm i)^\lambda\Gamma(\lambda )k^{-\lambda} (t\pm i\epsilon)^{-\lambda}
\over (\cosh y-\sinh y\cos\gamma(\theta,\theta_k))^\lambda}~~,
\lb{p13}
\ee
where the $i\epsilon$-prescription determines the way to go around
the $t=0$ singularity. With the help of this representation we can
obtain an explicit  decomposition of the asymptotic field
(\ref{p11}) in terms of data on the boundary of the light-cone.

Consider the region ${\cal A_+}$ in which the coordinate $t>0$.
With the help of (\ref{p13}) the asymptotic field (\ref{p11}) has
the form
\be
\phi(t,y,\theta)={1\over 2\pi i}\int_{d/2-i\infty}^{d/2+i\infty}
d\lambda ~t^{-\lambda}\int_{S^+_d}d\mu(\theta_k) \ G_{(\lambda)}(y,\theta,\theta_k)
\ \phi^{(d-\lambda)} (\theta_k)~~,
\lb{p14}
\ee
where
\be
G_{(\lambda)}(y,\theta,\theta_k)=
{g(\lambda)\over (\cosh y-\sinh y\cos (\theta,\theta_k))^\lambda}
\lb{p15}
\ee
is the bulk-boundary propagator on Euclidean anti-de Sitter
introduced in (\ref{14''}). The family of functions (parametrized
by $\lambda$)
\be
\phi^{(d-\lambda)} (\theta_k)={2^{d\over 2}\Gamma (\lambda-{d\over 2})\over 2\pi^{1\over 2}}
\int_0^\infty dk~k^{(d-\lambda-{1\over 2})}\left(e^{-i{\pi\over 2}\lambda} a(k,\theta_k)+
e^{i{\pi\over 2}\lambda} a^{\dagger}(k,\theta_k)\right)
\lb{p16}
\ee
can be thought of as the  boundary data on the boundary $S^+_d$ of
the future light-cone. Equation (\ref{p14}) thus gives us a
solution to a quite peculiar boundary problem and shows how the
field in (d+2)-dimensional space-time can be reconstructed from
data on a co-dimension two surface lying at the boundary at
infinity of the light-cone.

The algebra (\ref{p12}) induces the algebra of the operators
(\ref{p16}) and we obtain for their vacuum expectation values
\be
&&<0|\phi^{(d-\lambda)} (\theta_k)\phi^{(d-\lambda')} (\theta_k')|0>\nonumber \\
&&=
{2^d\over 4\pi}\Gamma (\lambda-{d\over 2})\Gamma (\lambda'-{d\over 2}) (-i)^\lambda i^{\lambda'}
2\pi i ~\delta (\lambda+\lambda'-d)~\delta^{(d)}(\theta_k,\theta_k')~~.
\lb{p17}
\ee
Using this we can compute the vacuum expectation value of the
asymptotic fields (\ref{p11}) and we find that
\be
&&<0|\{\phi (t,y,\theta ),\phi (t',y',\theta')\}|0> \nonumber \\
&&=
{1\over 2\pi i}\int_{d/2-i\infty}^{d/2+i\infty} d\lambda ~t^{-\lambda}~ t'^{-(d-\lambda)}~
 C(\lambda )\int_{S^+_d}d\mu(\theta_k) \ G_{(\lambda )}(y,\theta_k,\theta ) \
G_{(d-\lambda )}(y',\theta_k,\theta')~~,\nonumber  \\
&&C(\lambda )={2^d\over 2\pi}\ \Gamma (\lambda-{d\over 2}) \ \Gamma ({d\over 2}-\lambda) \ \cos ({\pi\over 2}(d-2\lambda))~~.
\lb{p18}
\ee
Equation (\ref{p18}) is another representation for the Hadamard
function (\ref{p6}). Comparing these  two expressions we find a
representation for the anti-de Sitter bulk-bulk propagator
$D_\lambda(\cosh\sigma )$ in terms of the bulk-boundary propagator
(\ref{p15}). This representation is rather natural from the point
of view of anti-de Sitter space.

In the region ${\cal A}_-$ where the coordinate $t<0$ there exists
a representation for the asymptotic field which is similar to
(\ref{p14}): one should just switch signs appropriately according
to the $i\epsilon$-prescription in (\ref{p13}).

In region $\cal D$ foliated by de Sitter slices the coordinates
$(X_0,\bf{X})$ are given by (\ref{X}) and we have that
$$
(kX_0-{\bf{k}}{\bf{X}})=kr~
(\sinh \tau-\cosh \tau\cos\gamma(\theta,\theta_k))~~.
$$
The expression for the plane waves then reads
\be
e^{\pm i(kX_0-{\bf{k}}{\bf{X}})}={1\over 2\pi i}\int_{c-i\infty}^{c+i\infty}
d \lambda {(\pm i)^\lambda\Gamma(\lambda )k^{-\lambda} r^{-\lambda}
\over
(\sinh \tau -\cosh \tau\cos\gamma(\theta,\theta_k) \pm i\epsilon)^\lambda}~~.
\lb{p19}
\ee
The kernel appearing in (\ref{p19}) is singular and ambiguous. The
$i\epsilon$ in (\ref{p19}) provides the appropriate way to deal
with the singularity in the kernel. The asymptotic behavior of
such kernels is given by (\ref{Glambda}) and (\ref{Glambda1}).
Using those asymptotics we define the bulk to boundary propagator
near the boundary at future infinity of the light-cone as follows
\be
G^{\tt out}_{\lambda }(\tau,\theta,\theta')={g(\lambda) ~(-i)^{d-2\lambda}\over
(\sinh \tau -\cosh \tau\cos\gamma(\theta,\theta' ) +i\epsilon)^\lambda}~~.
\lb{Gkern}
\ee
The bulk-boundary propagator near the boundary at past infinity is
defined as
\be
G^{\tt in}_\lambda (\tau,\theta,\theta')=
{g(\lambda)~ (-i)^{d}\over
(\sinh \tau +\cosh \tau\cos\gamma(\theta,\theta' ) -i\epsilon)^\lambda}~~.
\lb{Gkernin}
\ee
We should note that the $\epsilon$-prescription used to define the
decomposition of the plane waves (see (\ref{p13}) and (\ref{p19}))
is in general not analytic under the transformation (\ref{p7})
between regions $\cal A$ and $\cal D$. We find that modes
$e^{-ikX}$ with $k^0>0$ are analytic across the future light-cone
while modes $e^{ikX}$ with $k^0>0$ are analytic across the past
light-cone.

\subsection{Conformal operators}

The analysis of the
previous subsection suggests that  one can define the conformal operators
as associated with the asymptotic (possibly, interacting in the bulk)
in- and out- quantum fields. For the out-field the representation is
\be
\lim_{\tau\to+\infty}\phi(r,\tau,\theta)={1\over 2\pi i}\int_{{d\over 2}-i\infty}^{{d\over 2}+i\infty}
d\lambda \left( \varphi^<_\lambda (r,\tau) ~
{_{\tt out}}\!{\cal O}^<_\lambda (\theta)
+\varphi^>_\lambda (r,\tau)~{_{\tt out}}\!{\cal O}^>_\lambda (\theta) \right)~~.
\lb{p20}
\ee
For the in-field we have a similar representation in
terms of the in-operators
\be
\lim_{\tau\to-\infty}\phi(r,\tau,\theta)={1\over 2\pi i}\int_{{d\over 2}-i\infty}^{{d\over 2}+i\infty}
d\lambda \left(\chi^>_\lambda (r,\tau)~{_{\tt in}}\!{\cal O}^>_\lambda (\theta)
+\chi^<_\lambda (r,\tau)~{_{\tt in}}\!{\cal O}^<_\lambda (\theta)\right)~~.
\lb{p21}
\ee
The asymptotic modes which are right-moving as a function of
$\lambda$ on the ($\log r,\tau $)-plane are labelled by the
subscript $>$ while the left-moving modes are labelled by $<$. The
functions
\be
\varphi^<_\lambda (r,\tau)=N^{-1}_\lambda ~r^{-\lambda}e^{-\lambda\tau}~,~~
\varphi^>_\lambda (r,\tau)=N^{-1}_\lambda~i^{\lambda-d}~ r^{-\lambda}e^{-(d-\lambda)\tau}
\lb{modes-out}
\ee
form the basis of out-modes and the functions
\be
\chi^>_\lambda (r,\tau)=N^{-1}_\lambda ~r^{-\lambda}e^{\lambda\tau}~,~~
\chi^<_\lambda (r,\tau)=N^{-1}_\lambda ~i^{\lambda-d}~r^{-\lambda}e^{(d-\lambda)\tau}
\lb{modes-in}
\ee
form the basis of the in-modes on the $(\log r,\tau)$-plane. In
the region $\cal A$ the respective modes are those obtained from
(\ref{modes-out}) and (\ref{modes-in}) by means of the analytic
continuation (\ref{p7}). The conformal operators so defined are
associated with the boundary of the light-cone at infinity and are
independent of the side ($\cal A$ or $\cal D$) from which the
boundary is approached.

Since $\tau$ plays the role of time in the de Sitter foliation of
Minkowski space it can be used  to define positive and negative
frequencies. Representing $\lambda={d\over 2}+i\alpha$ where
$\alpha$ is a real parameter we then find that the positive
frequency modes are $\chi^>_{(-\alpha)}$, $\chi^<_{(\alpha)}$,
$\varphi^<_{(\alpha)}$ and $\varphi^>_{(-\alpha)}$, all with
$\alpha$ positive. Inversion of the sign of $\alpha$ (or
equivalently taking the complex conjugate) then gives the
negative-frequency modes. The constant $N_\lambda$ is given by
$$
N_\lambda=2^{-{d\over2}} \sqrt{2\lambda-d}~~.
$$
The positive frequency modes then are normalized as
\be
({\it f}_\lambda,{\it f}_{\lambda'})=2\pi i~ \delta(\lambda-\lambda')~~,
\lb{norm}
\ee
where $f_\lambda$ stands for any of the positive frequency modes
we defined above. The inner product is given by the usual
Klein-Gordon expression
\be
({\it f},{\it g})=(\cosh \tau)^d\int_0^\infty dr \ r^{d-1}{\it f}^*
\stackrel{\longleftrightarrow}{i\partial_\tau} {\it g}~~,
\lb{product}
\ee
where $\tau\rightarrow+\infty$ for the out-operators and
$\tau\rightarrow-\infty$ for in-operators.

For the operators defined in (\ref{p20}) and (\ref{p21})  we have
that ${\cal O}^<_{-\alpha}=[{\cal O}^<_{\alpha}]^*$ and ${\cal
O}^>_{-\alpha}=[{\cal O}^>_{\alpha}]^*$. This implies that the in-
and out-operators ${\cal O}^<_\alpha$ and  ${\cal O}^>_{-\alpha}$
couple to positive-frequency modes while ${\cal O}^>_\alpha$ and
${\cal O}^<_{-\alpha}$ couple to negative-frequency modes. This
definition of the positive and negative frequencies is different
from the definition based on the global time coordinate $X_0$. The
difference between two is  reminiscent of the well-known relation
between the Rindler and Minkowski definitions of time in Rindler
spacetime. The corresponding modes are related by a Bogoliubov
transform. As is seen from (\ref{p19}) the modes which have
positive frequency in the sense of the time-like vector
$\partial_{X_0}$ decompose onto both negative and positive
frequency modes in the sense of de Sitter time vector
$\partial_\tau$. It would be interesting to understand better the
implications of this observation. Note also that the definition of
the conformal operators given in (\ref{p20}) and (\ref{p21}) is
reminiscent of the analysis of section~4.

The operation inverse to the  spectral integration in (\ref{p20}),
(\ref{p21}) is the direct Mellin transform
$$
\phi_\lambda (\tau,\theta )=\int^\infty_0dr \ r^{\lambda-1}\phi (r,\tau, \theta )
$$
so that $\phi_\lambda(\tau,\theta)$ can be thought of as a field
living on de Sitter space with coordinates $(\tau,\theta )$. On
the other hand the kernels $G^{\tt in}_\lambda
(\tau,\theta,\theta')$ and $G^{\tt out }_\lambda
(\tau,\theta,\theta')$ given in (\ref{Gkern}) and (\ref{Gkernin})
can be used to project\footnote{ There are two terms in the
asymptotic expansion of $G^{\tt in (out)}_\lambda
(\tau,\theta,\theta')$, see (\ref{Glambda}) and (\ref{Glambda1}).
In order for the term with the delta-function to dominate we take
$Re~\lambda={d\over 2}+\epsilon$ and take $\epsilon$ to zero
afterwards.} the field $\phi_\lambda (\tau,\theta )$ onto modes
asymptotic to $e^{\pm\lambda\tau}$ and $e^{\pm(d-\lambda)\tau}$.
These two operations can be used to derive the insertion of the
in- and out-conformal operators in terms of the (in general
interacting) bulk Feynman  Green's function in the form of an
integral over a hypersurface of constant $\tau$. For the
out-operators it is the surface at $\tau=+\infty$ while for the
in-operators  we should take the surface at $\tau=-\infty$. The
insertion prescriptions are
\be
&&<0|..~ {_{\tt in}}\!{\cal O}^<_\lambda (\theta)~..|0>= \\
&&\lim_{\tau\to -\infty}
\int_0^\infty dr r^{\lambda-1} N^{-1}_\lambda ~i^{d-\lambda}\int_{S^-_d} d\mu(\theta') (\cosh\tau )^d
D_{\tt F}(..,(r,\tau,\theta'),..)
\stackrel{\longleftrightarrow}{\partial_\tau} G^{\tt in}_{d-\lambda} (\tau,\theta,\theta')\nonumber 
\lb{p22}
\ee
and
\be
&&<0|..~ {_{\tt in}}\!{\cal O}^>_\lambda (\theta)~..|0>=  \\
&&\lim_{\tau\to -\infty}
\int_0^\infty dr r^{\lambda-1} N^{-1}_\lambda \int_{S^-_d} d\mu(\theta') (\cosh\tau )^d
D_{\tt F}(..,(r,\tau,\theta'),..)
\stackrel{\longleftrightarrow}{\partial_\tau} G^{\tt in}_{\lambda} (\tau,\theta,\theta')\nonumber 
\lb{p23}
\ee
for the in-operators and
\be
&&<0|..~ {_{\tt out}}\!{\cal O}^<_\lambda (\theta)~..|0>=  \\
&&\lim_{\tau\to +\infty}
\int_0^\infty dr r^{\lambda-1} N^{-1}_\lambda \int_{S^+_d} d\mu(\theta') (\cosh\tau )^d
G^{\tt out}_{\lambda} (\tau,\theta,\theta')\stackrel{\longleftrightarrow}{\partial_\tau}
D_{\tt F}(..,(r,\tau,\theta'),..)\nonumber 
\lb{p24}
\ee
and
\be
&&<0|..~ {_{\tt out}}\!{\cal O}^>_\lambda (\theta)~..|0>= \\
&&\lim_{\tau\to +\infty}
\int_0^\infty dr r^{\lambda-1} N^{-1}_\lambda ~i^{d-\lambda}\int_{S^+_d} d\mu(\theta') (\cosh\tau )^d
G^{\tt out}_{d-\lambda} (\tau,\theta,\theta')
\stackrel{\longleftrightarrow}{\partial_\tau}
D_{\tt F}(..,(r,\tau,\theta'),..)\nonumber 
\lb{p25}
\ee
for the out-operators. Near the boundary ($S^+$ or $S^-$) of the
light-cone the coordinate $r$ plays the role of a coordinate
enumerating the de Sitter slices ending at the surface $S^+$ (or
$S^-$). Thus, in the above formulas an insertion effectively takes
place at each of the de Sitter slices separately, after which one
sums over all de Sitter slices labelled by $r$.

These expressions are very similar to the ones introduced in
\cite{Str,Spradlin:2001nb} for the insertion of boundary conformal
operators in the case of de Sitter space. The field $\phi_\lambda
(\tau,\theta )$ obtained from $\phi (r,\tau,\theta )$ by the
Mellin transform in the variable $r$ describes a field on de
Sitter space with mass equal $\lambda (d-\lambda)$, so perhaps
this is not too surprising. Nevertheless, it is interesting to
observe that computing the on-shell action (as we attempted in
section~4) is not very satisfactory, whereas the S-matrix approach
in this section gives a concise proposal for a Minkowski/CFT
duality as in (\ref{p25}), just as in de Sitter space.

Inside the light-cone, in the region $\cal A$, the insertions can
be defined in a similar way as in (\ref{p22}), (\ref{p23}),
(\ref{p24}), (\ref{p25}). The appropriate surface is then the
boundary at $y=+\infty$ of each anti-de Sitter slice and one has
to integrate over contributions of all such slices, which
translates into an integral over negative $t$ for the $\tt
in$-operators and over positive $t$ for the $\tt out$-operators.
On an anti-de Sitter slice we could use the definition of the
conformal correlation functions built out of bulk-boundary
propagators and truncated n-point Feynman Green's function on the
slice, as familiar from AdS/CFT, see \cite{Wit},
\cite{Balasubramanian:1999ri}. Notice that even on the Anti-de
Sitter slices computing an on-shell action is not the right
approach, because of the complex values of the conformal weights,
and therefore even on the AdS slices we should resort to an
S-matrix type approach like the one discussed in
\cite{Giddings:1999qu}.

The integration over $t$ adds up the contributions of all AdS
slices. The explicit form of the insertions in region $\cal A$ can
be obtained by applying the analytic transformation (\ref{p7}) to
(\ref{p22})-(\ref{p25}).

It appears that correlation functions defined in this way do not
depend on the choice of region (inside or outside the light-cone)
which is used to define the insertions and are therefore an
intrinsic property of the boundary of the light-cone. The 2-point
correlation functions between $\tt out$- and $\tt in$-operators
are now found to be equal to
\be
<0| {_{\tt out}}\!{\cal O}^<_{\lambda_1} (\theta_1)~
{_{\tt in}}\!{\cal O}^>_{\lambda_2} (\theta_2)|0> = 2\pi i~\delta (\lambda_1+\lambda_2-d)~\delta^{(d)}(\theta_1,\pi-\theta_2)~i^d N_{\lambda_1} N_{\lambda_2}~F(\lambda_2)
\lb{p26}
\ee
\be
<0| {_{\tt out}}\!{\cal O}^>_{\lambda_1} (\theta_1)~
{_{\tt in}}\!{\cal O}^<_{\lambda_2} (\theta_2)|0> = 2\pi i~\delta (\lambda_1+\lambda_2-d)~\delta^{(d)}(\theta_1,\pi- \theta_2)~i^d N_{\lambda_1} N_{\lambda_2}~F(\lambda_1)
\lb{p27}
\ee
\be
<0| {_{\tt out}}\!{\cal O}^>_{\lambda_1} (\theta_1)~
{_{\tt in}}\!{\cal O}^>_{\lambda_2} (\theta_2)|0> = 2\pi i~\delta (\lambda_1+\lambda_2-d)~{i^{d-\lambda_2} N_{\lambda_1} N_{\lambda_2}~D(\lambda_2)\over (1+\cos\gamma(\theta_1,\theta_2))^{\lambda_2}}
\lb{p28}
\ee
\be
<0| {_{\tt out}}\!{\cal O}^<_{\lambda_1} (\theta_1)~
{_{\tt in}}\!{\cal O}^<_{\lambda_2} (\theta_2)|0> = 2\pi i~\delta (\lambda_1+\lambda_2-d)~{i^{d-\lambda_1} N_{\lambda_1} N_{\lambda_2}~D(\lambda_1)\over (1+\cos\gamma(\theta_1,\theta_2))^{\lambda_1}}~~.
\lb{p29}
\ee
Thus, the set of CFT operators splits into two groups: operators
${\cal O}^>$ representing the right-moving modes and operators
${\cal O}^<$ representing the left-moving modes. The correlation
functions of operators of the same group take a standard CFT form
(\ref{p28}), (\ref{p29}) on a d-sphere. On the other hand,
operators from different groups produce contact terms in the
correlators (\ref{p26}), (\ref{p27}). Note that contact type
correlation functions are also consistent with conformal
invariance.

\subsection{S-matrix}

It is well-known that in Minkowski space the S-matrix of
interacting quantum fields can be reconstructed from bulk
correlation functions via the so-called LSZ construction. A
natural question arises whether it is possible to reproduce the
S-matrix by means of only the light-cone correlation functions of
conformal operators introduced in section 5.3. That this may be
the case is suggested by the analysis made in de Sitter space
\cite{Spradlin:2001nb} where it was shown that the de Sitter
analog of an S-matrix indeed can be presented in terms of CFT
correlation functions between operators living at the infinite
past and future of de Sitter space. In anti-de Sitter a similar
construction was introduced in \cite{Giddings:1999qu}.

In order to analyze this possibility we consider the one-particle
amplitude whose LSZ form is given by a combination of integrating
over two surfaces, one lying in the infinite past and another in
the infinite future of Minkowski space \cite{BjorkenDrell},
\be
&&<p|q>=\int {dV(X)\over \sqrt{Z}} \int {dV(Y)\over \sqrt{Z}} f_q(X) \stackrel{\longrightarrow}{\nabla^2}_X D_{\tt F}(Y,X)\stackrel{\longleftarrow}{\nabla^2}_Y
f^*_p(Y)\nonumber \\
&&=-\int_{Y_0\to+\infty}d\sigma(Y)\int_{X_0\to-\infty}d\sigma(X)~ n^\mu_x~
f_q(X)\stackrel{\longleftrightarrow}{\partial_\mu} D_{\tt F}(Y,X)\stackrel{\longleftrightarrow}
{\partial_\nu}f^*_p(Y)~n^\nu_y~~,
\lb{p30}
\ee
where $Z$ is the standard normalization of the wave function and
$n^\mu_x$ is the normal vector to the hypersurface in the infinite
past while $n^\mu_y$ is the normal to the hypersurface in the
infinite future. The functions $f_k(X)$ are defined in (\ref{fk})
and form a basis of the set of asymptotic fields. We recall that
in momentum space the asymptotic spatial momentum $\bf{k}$ is
determined by its length $k$ and angles $\{\theta_k\}$ on a
d-sphere so that ${\bf{k}}=k ~{\bf{n}}(\theta_k)$.

\begin{figure}
\centerline{\psfig{figure=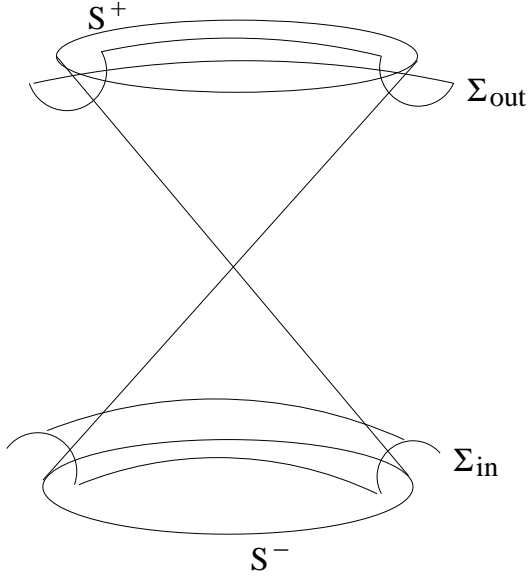}} \caption{The asymptotic
hypersurfaces $\Sigma_{\tt in}$ and $\Sigma_{\tt out}$ are defined
in a small vicinity of the past and future infinity of the
light-cone respectively.}
\end{figure}
For our purposes it is convenient to choose the asymptotic
hypersurfaces in a different way. Namely, we define the asymptotic
future and past hypersurfaces by $ \Sigma_{\tt out}=\{{\cal
A}:~y={\rm const}\to\infty~(t>0)\} \cup \{{\cal D}:~\tau={\rm
const} \to\infty \} $ and $\Sigma_{\tt in}=\{{\cal A}:~y={\rm
const}\to \infty~(t<0)\} \cup \{{\cal D}:~\tau={\rm const} \to
-\infty \}~~. $ We see that these hypersurfaces are defined in the
vicinity of the boundary of the light-cone and consist of two
components: one in region $\cal A$ and another in region $\cal D$.
As we show in Appendix \ref{normalization} the plane waves
$f_q(X)$ have the standard normalization with respect to the
Klein-Gordon inner product defined at these hypersurfaces. Thus,
$\Sigma_{\tt out}$ and $\Sigma_{\tt in}$ can be used to define
asymptotic states in the same fashion as the usual constant $X_0$
surfaces. With this definitions and these choices of asymptotic
surfaces the LSZ form of the one-particle amplitude (\ref{p30}) is
modified into
\be
<p|q>=\int_{\Sigma_{\tt out}} {d\sigma \over \sqrt{Z}}
\int_{\Sigma_{\tt in}} {d\sigma' \over \sqrt{Z}}
f^*_p (X) \stackrel{\longleftrightarrow}{\partial_n} D_F (X,X')
\stackrel{\longleftrightarrow}{\partial_n'} f_q(X')~~.
\lb{pq}
\ee
The asymptotic functions $f_q(X)$ and $f^*_p(Y)$ can be decomposed
according to (\ref{p19}) in the region $\cal D$ and  (\ref{p13})
in the region $\cal A$. Inserting  this decomposition into
(\ref{pq}) and after some manipulations we recognize the two-point
functions defined in section 5.3. Consequently, the one-particle
amplitude takes the form
\be
&&<p|q>= {\sqrt{p~q}\over (2\pi i )^2}
\int {d\lambda_1\over \sqrt{Z}}\int {d\lambda_2\over \sqrt{Z}}
~p^{-\lambda_1}q^{-\lambda_2}~H(\lambda_1)~H(\lambda_2)~\nonumber \\
&&<0| {_{\tt out}}\!{\cal O}^<_{d-\lambda_1} (\theta_p)~
{_{\tt in}}\!{\cal O}^>_{d-\lambda_2} (\pi-\theta_q)|0> ~~,
\lb{p32}
\ee
where
$$
H(\lambda)={2^{d\over 2}\over\sqrt{\pi}}~i~\Gamma(\lambda-{d\over 2})~\sin {\pi\over 2}(d-2\lambda )
$$
and each $\lambda$-integration is from ${d\over 2}-i\infty$ to
${d\over 2}+i\infty$. The function $f^*_p(X)$, once expressed in
terms of conformal operators, is not analytic across the future
light-cone. The jump is $i^{d-2\lambda}-i^{2\lambda-d}$, which
explains the appearance of $\sin{\pi\over 2}(d-2\lambda )$ in the
function $H(\lambda)$. The same is true for the function $f_q(X)$
near the boundary at infinity of the past light-cone. For free
fields the light-cone 2-point function in (\ref{p32}) is given by
(\ref{p26}) and after the integration over $\lambda_1$ and
$\lambda_2$ the usual expression
\be
<p|q>~=~ p^{-d}~\delta(p-q)~\delta^{(d)}(\theta_p,\theta_q)
\lb{amplitude}
\ee
is reproduced.  Although this calculation was performed for a free
field interaction is not expected to change the form of the
conformal correlation function in the amplitude (\ref{p32}), but
only its normalization. Therefore the angle part of the amplitude
comes out in the same way as in (\ref{amplitude}) while the
$p$-dependent pre-factor can be modified. Note that this is also
consistent with the restrictions imposed by bulk Lorentz
invariance on the structure of the one-particle amplitude.

The generalization of (\ref{p32}) to many-particle amplitudes is
straightforward. It is important to note that the amplitude
(\ref{p32}) is given in terms of the contact type correlation
functions of operators introduced in section~5.3.

In the construction of the scattering amplitudes (\ref{p30}) and
(\ref{p32}) the positive and negative frequency modes of the
asymptotic fields are defined with respect to the vector field
$\partial_{X_0}$ which is a global time-like vector in Minkowski
space. Instead we could have chosen to use the time $\tau$ on a de
Sitter slice to define the positive and negative frequency modes.
The basis of asymptotic fields then is given by the functions
(\ref{modes-in}) and (\ref{modes-out}). These modes can be used to
form incoming ($f$) and outgoing $(g)$ wavepackets as
\be
f^>(r,\tau,\theta)={1\over 2\pi i}\int d\lambda ~
\chi^>_\lambda(r,\tau)~f^>_\lambda (\theta)~,~~
f^<(r,\tau,\theta)={1\over 2\pi i}\int d\lambda ~
\chi^<_\lambda(r,\tau)~f^<_\lambda (\theta)~, \nonumber \\
\noindent g^>(r,\tau,\theta)={1\over 2\pi i}\int d\lambda ~
\varphi^>_\lambda(r,\tau)~g^>_\lambda (\theta)~,~~
g^<(r,\tau,\theta)={1\over 2\pi i}\int d\lambda ~
\varphi^<_\lambda(r,\tau)~g^<_\lambda (\theta)~,
\lb{p33}
\ee
where $f^>_\lambda (\theta)$ ($f^<_\lambda (\theta)$) and
$g^>_\lambda (\theta)$ ($g^<_\lambda (\theta)$) are smooth
functions on the spheres $S^-_d$ and $S^+_d$ respectively. In
order to make sure that only positive frequency modes contribute
to the wavepacket we demand that (provided that $\lambda={d\over
2}+i\alpha$) $f^>_{\alpha} (\theta)=0$ and $f^<_{-\alpha}(\theta
)=0$ for $\alpha>0$, with similar conditions on the envelope
functions in the outgoing wavepacket. The operators $b^{>(<)}_f$
and $b^{>(<)}_g$ corresponding to these wavepackets are recovered
from the asymptotic free field (\ref{p20}) (or (\ref{p21})) by the
standard formulas
\be
b^{>}_f=\int_0^\infty dr~r^{d-1}\int_{S^-_d}d\mu (\theta) ~(\cosh\tau)^d~ [f^{>}(r,\tau,\theta)]^*
i\stackrel{\longleftrightarrow}{\partial_\tau} \phi_{\tt in} (r,\tau,\theta)
\nonumber \\
b^{>}_g=\int_0^\infty dr~r^{d-1}\int_{S^+_d}d\mu (\theta) ~(\cosh\tau)^d~ [g^{>}(r,\tau,\theta)]^*
i\stackrel{\longleftrightarrow}{\partial_\tau} \phi_{\tt out} (r,\tau,\theta)
\lb{p34}
\ee
with similar expressions for the left-moving modes. The S-matrix
elements for an incoming wavepacket $f^>$ and an outcoming
wavepacket $g^>$ are defined by
\be
S[g^>,f^>]=<0|b^>_g~b^{>\dagger}_f|0>~~.
\lb{p35}
\ee
These matrix elements can be re-written in terms of 2-point
correlation functions of the conformal operators ${\cal O}^>$ as
\be
S[g^>,f^>]= {1\over (2\pi i)^2}\int {d\lambda \over \sqrt{Z}}
\int {d\lambda'\over \sqrt{Z}} \int_{S^+_d}
[g^>_\lambda (\theta)]^* \int_{S^-_d}f^>_{\lambda'}(\theta')
<0| {_{\tt out}}\!{\cal O}^>_{d-\lambda} (\theta)
{_{\tt in}}\!{\cal O}^>_{\lambda'} (\theta')|0> ~~
\label{p36}
\ee
A similar expression exists for the matrix element $S[g^<,f^<]$.
This formula relates the matrix elements of the S-matrix with the
light-cone conformal correlation functions of operators living on
the asymptotic d-spheres. A generalization of this formula to
many-particle amplitudes is straightforward. Our analysis suggests
that the S-matrix approach to Quantum Field Theory in Minkowski
space can be re-formulated entirely in terms of the correlation
functions of conformal operators living on the boundaries of the
light-cone. This opens the possibility to find a truly holographic
representation of the physics in Minkowski space. For such a
holographic representation we should also consider, obviously,
gravity in Minkowski space, to which we turn next.

\
\section{Asymptotic light-cone structure and \\
conformal  symmetry}
\setcounter{equation}0

\subsection{Asymptotic Minkowski Spaces}

In this section we consider a more general class of metrics which
are not globally Minkowski but approach flat space in a suitable
asymptotic sense. In particular, we are interested in the
asymptotic light-cone structure of the metric and the asymptotic
symmetries associated to this structure.

First, we generalize the Minkowski metric studied in the previous
sections. It should approach the Minkowski metric asymptotically
in such a way that at infinity a slicing with constant curvature
surfaces is appropriate. For concreteness we consider only the
region ${\cal A}_-$.  We start with the following asymptotic form
of the (d+2)-metric
\be
ds^2=-dt^2+t^2\left({d\rho^2\over 4\rho^2}(1+t^{-1}\sigma(t,\theta,\rho ))+
{1\over \rho}g_{ij}(t, \theta,\rho)d\theta^id\theta^j+2t^{-1}A_i(t,\theta , \rho)
d\theta^i d\rho
\right)
\lb{6.1}
\ee
where normal coordinates, in which the components $G_{t\rho}$ and
$G_{ti}$ vanish, have been used. The metric in parenthesis in
(\ref{6.1}) is the $(d+1)$-metric on an Euclidean slice of
constant $t$. Provided that the $(d+2)$-metric is Ricci flat, the
$t={\rm const}$ slice of the light-cone metric (\ref{6.1}) should
for large $t$ approach a solution of the Einstein equations with
negative cosmological constant. Therefore, the choice of the
metric (\ref{6.1}) to the leading order in $t$ is motivated by the
general form of the negative constant curvature metric studied in
the context of the AdS/CFT correspondence \cite{HS}. The conformal
boundary of a $t={\rm const}$ slice is at $\rho=0$. For large $t$ 
each function in (\ref{6.1}) is assumed to have an expansion in powers of $1/t$
as well as in powers of $\rho^{1/2}$. For instance, the metric
$g_{ij}(t,\theta ,\rho)$ can be expanded as
$$
g_{ij}(t,\theta ,\rho)=g^{(0)}_{ij}(\theta, \rho )+t^{-1}g^{(1)}_{ij}(\theta, \rho )
+...~~,
$$
which could also include logarithmic terms of the form $t^{-n}(\ln
t)^{-m}$ and each term in the $t$-expansion should be also
expanded in terms of $\rho^{1/2}$, e.g.
$$
g^{(n)}_{ij}(\theta, \rho )=g^{(n,0)}_{ij}(\theta )+g^{(n,1)}_{ij}(\theta )\rho^{1/2}
+...~~,
$$
where $g^{(0,0)}_{ij}(\theta )$ is to be interpreted as the metric
on the boundary $\Sigma^{-}_d$ of the past light-cone. 
This boundary metric should be the same for all $t={\rm const}$ slices. 
Expanding first in powers of $\rho$ this  condition says that the 
constant term in the expansion should not be a function of $t$,
i.e. $g_{ij}(t,\theta,\rho)=g^{(0,0)}_{ij}(\theta )+O(\rho^{1/2})$.
For large $t$ this gives a restriction on the coefficients 
$g^{(n,0)}_{ij}$, namely only the one with $n=0$ should be non-vanishing.
We do the holography along the $\rho$-direction so that no additional 
expansion in $t$-variable should  be necessary. It is however technically
more convenient to work with functions which have certain expansion in 
powers of $1/t$.
A bonus we get is extra diffeomorphisms which generate transformations
in the $t$ directions  and change the metric to $1/t$ order.
The BMS group will be a part of these diffeomorphisms.

When we analyze the bulk diffeomorphisms that preserve the
asymptotic form  (\ref{6.1}) we find that there is a particular
$t$-independent diffeomorphism generated by
\be
\xi^t=0~~,~~~\xi^\rho=\gamma (\theta )\rho~~, ~~\xi^k=\xi^k(\theta, \rho )~~,
\lb{diff}
\ee
where $\gamma (\theta )$ is some arbitrary function of $\theta $,
$\xi^k(\theta, \rho )$ has to satisfy
\be
\partial_\rho \xi^k=-{1\over 4}g^{ki}_{(0)}\partial_i \gamma (\theta )~~.
\lb{diff1}
\ee
The action of this diffeomorphism on  $\sigma (t,\theta , \rho )$,
$g_{ij}(t,\theta , \rho )$ and $A_i(t,\theta , \rho )$ is then
given  by
\be
{\cal L}_\xi \sigma=\gamma (\theta )\rho\partial_\rho \sigma
-2\rho^2A_k g^{ki}_{(0)}\partial_i\gamma +\xi^i\partial_i\sigma
\lb{sigma}
\ee

\be
{\cal L}_\xi g_{ij}=\gamma (\theta )(\rho\partial_\rho g_{ij}-g_{ij})
+{\rho^2\over t}(A_i\partial_j\gamma+A_j\partial_i\gamma)+\nabla^{(g)}_i\xi_j+
\nabla^{(g)}_j\xi_i
\lb{gij}
\ee

\be
{\cal L}_\xi A_i=\gamma (\theta )(\rho\partial_\rho A_i +A_i)+
{t\over 4\rho}(\partial_i\gamma-g_{ki}g^{kj}_{(0)}\partial_j\gamma)+\xi^k\partial_k A_i+
\partial_i\xi^k A_k
\lb{Ai}
\ee
Notice that although there appears to be a term (the one
proportional to $t$) in the right hand side of (\ref{Ai}) which
destroys its asymptotic behavior, this is not really true
because $\partial_i\gamma-g_{ki}g^{kj}_{(0)}\partial_j\gamma$
vanishes as $1/t$ in the limit where $t\rightarrow\infty$. One can
see from (\ref{gij}) that the diffeomorphism generated by
(\ref{diff}) acts on the metric $g^{(0)}_{ij}(\theta , \rho)$ in
the same way as the diffeomorphism that was found in \cite{ISTY}
to generate the asymptotic conformal structure of asymptotically
Anti-de Sitter space. In particular, it acts on the metric
$g^{(0,0)}_{ij}(\theta )$, the metric on the asymptotic boundary
$\Sigma$ of the light-cone,
 as a conformal transformation
\be
{\cal L}_\xi g^{(0,0)}_{ij}=-\gamma (\theta )g^{(0,0)}_{ij}~~.
\lb{conf}
\ee

Thus, our analysis demonstrates that the conformal structure
present on each AdS slice extends to the conformal structure
associated with the boundary of the light-cone of
$(d+2)$-dimensional asymptotically Minkowski space-time.

The analysis done so far was restricted to the region ${\cal
A}_-$. A similar analysis can be done in the regions ${\cal A}_+$
and $\cal D$. In fact the conformal structure near the boundary
$\Sigma^{-}_d$ of the past light-cone is analytic across the
light-cone and extends from region ${\cal A}_-$ to region ${\cal
D}$. This can be seen by introducing new light-cone coordinates
$u=-{t\over \rho^{1/2}}$, $v=t\rho^{1/2}$ so that
\be
uv=-t^2~~,~~~{v\over u}=-\rho~~.
\lb{6.4}
\ee
In terms of these coordinates the past light-cone is defined  by
$v=0$. The boundary $\Sigma^-_d$ of the light-cone is now at
$u=-\infty$. In terms of the new coordinates the metric
(\ref{6.1}) takes the light-cone form (for simplicity we discard
the subleading $1/t$-terms in  the metric (\ref{6.1}))
\be
ds^2=dudv+u^2 g^{(0)}_{ij}(\theta,v/u)d\theta^id\theta^j~~.
\lb{6.5}
\ee
The boundary of the light-cone is at $u=-\infty$, and the
$\rho$-expansion of the metric components now becomes an expansion
in  $v/u$. In the light-cone form (\ref{6.5}) the metric is
analytic across the past light-cone (at $v=0$) and extends from
${\cal A}_-$ to ${\cal D}$ everywhere in the neighborhood of the
boundary of the past-light cone.

The diffeomorphism (\ref{diff}) now takes the form
\be
\xi^v={1\over 2}v\gamma (\theta)~~, ~~\xi^u=-{1\over 2}u\gamma (\theta)~~, ~~\xi^i=\xi^i (u,v,\theta )~~.
\lb{6.11}
\ee
The form of the functions $\xi^i(\theta, u,v)$ is constrained by
the equations
\be
\partial_u\xi^i=-{v\over 4u^2}g^{ik}_{(0)}(\theta,v/u)\partial_k \gamma~,~~
\partial_v\xi^i={1\over 4u}g^{ik}_{(0)}(\theta,v/u)\partial_k \gamma~~.
\lb{diffeo}
\ee
Since these are two equations there exists a consistency condition
$\partial_u\partial_v\xi^i=\partial_v\partial_u\xi^i$. It reduces
to a condition $(v\partial_v+u\partial_u)g^{(0)}_{ij}=0$ which
does hold since the metric components in (\ref{6.5}) are functions
of $v/u$. One can see from (\ref{diffeo}) that $\xi^i$ is a
function of the ratio $v/u$ of the light-cone coordinates. The
first term $\xi^i_{(0)}$ in the expansion of $\xi^i(\theta, v/u)$
near the boundary of the light-cone is proportional to $v/ u$ and
solving (\ref{diffeo}) to this order we find that
$\xi^i_{(0)}={v\over 4u}g_{(0,0)}^{ik}\partial_k \gamma (\theta)$.

The components $g^{(0)}_{ij}(\theta,v/u)$ transform under these
diffeomorphisms according to
\be
{\cal L}_\xi g^{(0)}_{ij}=\gamma (\theta )(v\partial_v g^{(0)}_{ij}-u^{-1}
\partial_u(u^2g^{(0)}_{ij}))+\nabla^{(0)}_i\xi_j+\nabla^{(0)}_j\xi_i~~.
\lb{mtransf}
\ee
Of course, equations (\ref{6.11}), (\ref{diffeo}) and
(\ref{mtransf}) are just the light-cone form of the previous
equations (\ref{diff}), (\ref{diff1}) and (\ref{gij}) that were
written in terms of the coordinates $(t,\rho )$. However, the
analyticity of the conformal structure across the light-cone is
not obvious in those coordinates.

We conclude that there indeed exists a conformal symmetry
associated with the boundary of the light-cone in the metric
(\ref{6.1}) or (\ref{6.5}). This symmetry is generated by a
certain bulk diffeomorphism (\ref{6.11}), (\ref{diffeo}) and can
be viewed as an extension of the conformal group existing on each
EAdS slice to a symmetry of the bulk.

\subsection{BMS}

It should be noted that the group of all diffeomorphisms
preserving the structure of the metric (\ref{6.1}) is larger than
just (\ref{diff}) and includes also $t$-dependent diffeomorphisms.
The  full analysis of diffeomorphisms of this type is given in
Appendix~C. We also find the generators of the well-known BMS
group as a particular case of these $t$-dependent asymptotic
transformations. The vector generating these diffeomorphisms is
\be
\xi^t=f(\theta )\rho^{1/2} ~~, \nonumber \\
\xi^\rho=-{2\over t}f(\theta) \rho^{3/2}~~, \nonumber \\
\xi^i={1\over t}\rho^{3/2}~g_{(0,0)}^{ij}(\theta)\partial_j f(\theta)~~.
\label{BMS1}
\ee
The BMS group  acts at a subleading order in the $1/t$ expansion
and leaves the metric $g^{(0,0)}_{ij}(\theta)$ on $\Sigma_d$
unchanged. The BMS transformations are the asymptotic symmetries
of the class of metrics which approach Minkowski metric near null
infinity. Our asymptotic condition (\ref{6.1}) is less restrictive
and requires only that there is a light-cone structure near the
surface $\Sigma_d$, which can be viewed as a single point in the
null infinity. Therefore the group of asymptotic symmetries in our
case is much larger than just BMS and in fact contains it as a
subgroup. Of course, one may wonder whether the group of
asymptotic symmetries and in particular the BMS group is actually
a symmetry of the dual theory. To analyze this issue we now turn
to a study of the various symmetries of the system.

\subsection{Symmetries of the dual theory}

First we recall that a solution of the field equations of a
massless scalar field admits on the de Sitter slices near
$\tau=+\infty$ a decomposition given in (\ref{p20})
\be
\lim_{\tau\to+\infty}\phi(r,\tau,\theta)={1\over 2\pi
i}\int_{{d\over 2}-i\infty}^{{d\over 2}+i\infty} d\lambda
N_{\lambda}^{-1} r^{-\lambda} \left( e^{-\lambda\tau} ~ {_{\tt
out}}\!{\cal O}^<_\lambda (\theta) + i^{\lambda-d}
e^{-(d-\lambda)\tau} ~{_{\tt out}}\!{\cal O}^>_\lambda (\theta)
\right). \label{jj1}
\ee
We have been applying holography in the $\tau$-direction,
resulting in the infinite set of operators ${_{\tt out}}\!{\cal
O}^<_\lambda (\theta)$ and ${_{\tt out}}\!{\cal O}^>_\lambda
(\theta)$. Instead of working with an infinite set of operators
parametrized by the continuous parameter $\lambda$, we could also
have decided replace the continuous parameter by an extra
coordinate, so that the dual theory seems to live in $d+1$
dimensions. One way to introduce such an extra coordinate is to
define
\be
\phi^{<}(\xi,\theta) ={1\over 2\pi i}\int_{{d\over
2}-i\infty}^{{d\over 2}+i\infty} d\lambda N_{\lambda}^{-1}
\xi^{-\lambda} {_{\tt out}}\!{\cal O}^<_\lambda (\theta) .
\ee
and similarly
\be
\phi^{>}(\xi,\theta) ={1\over 2\pi i}\int_{{d\over
2}-i\infty}^{{d\over 2}+i\infty} d\lambda N_{\lambda}^{-1}
\xi^{-\lambda} i^{\lambda-d}~{_{\tt out}}\!{\cal O}^>_\lambda
(\theta) .
\ee
In other words, the new coordinate $\xi$ arises by doing an
inverse Mellin transform over the parameter $\lambda$. In
constructing $\phi^{<,>}(\xi,\theta)$ we essentially did nothing
but drop the leading exponential behavior in $\tau$ in
(\ref{jj1}). Thus $\phi^{<,>}(\xi,\theta)$ describe the boundary
behavior of $\phi$ as $\tau\rightarrow \infty$ and in that sense
it is similar to how one defines dual fields in AdS.

We have on purpose called the extra coordinate $\xi$. It clearly
is very similar to $r$, but to make clear that we really want to
view it as a new coordinate we gave it a new name. It is actually
also closely related to $x_+$ that appears in (\ref{coorD}).
Indeed, $r^{-\lambda} e^{-\lambda\tau}=(x_+)^{-\lambda}$, and
therefore $\phi^<(\xi,\theta)$ describes the field $\phi$ as a
function of $x^+$ if we identify $x_+$ with $\xi$. From this
perspective it is also tempting to view the fields
$\phi^{<,>}(\xi,\theta)$ as fields living on null infinity, and to
view $\xi$ as the affine null coordinate of null infinity, so that
the dual theory would actually live on null infinity. As we will
see, this interpretation is somewhat problematic, but it would be
interesting to study this further.
On the other hand, $r^{-\lambda}e^{-(d-\lambda)\tau}=
(x_+)^{-d/2}(-x_-)^{d/2-\lambda}$ and dropping the overall factor
$(x_+)^{-d/2}$ the operator $\phi^>(\xi,\theta)$ can be considered as function
of $x_-$. Since $x_-$ is the affine coordinate along the light-cone
this suggests yet another interpretation of $\xi$ and of the dual theory.

We will now first study the action of the conformal group, or
equivalently, the Lorentz group $SO(d+1,1)$. To write the explicit
form of these generators in the de Sitter coordinates
$(\tau,r,\theta)$ it is convenient to combine the angles $\theta$
and $r$ in a $d+1$ component vector with components $r_i$, so that
$r^2 = \sum_i r_i^2$, and so that the angles parametrize the unit
sphere at $r=1$. With this convention, the rotation and boost
generators of Minkowski space are
\begin{eqnarray}
M_{ij} & = & r_i\frac{\partial}{\partial r_j} - r_j
 \frac{\partial}{\partial r_i} \nonumber \\
 K_i & = & \frac{r_i}{r} \frac{\partial}{\partial \tau} +
 \tanh \tau \left( r \frac{\partial}{\partial r_i} - \frac{r_i}{r}
 \sum_k r_k \frac{\partial}{\partial r_k}\right) . \label{jj2}
 \end{eqnarray}
In order to determine the action of these generators on
$\phi^{<,>}(\xi,\theta)$, the strategy is to first find the exact
solution of the free field equation with asymptotic behavior
determined by $\phi^{<,>}(\xi,\theta)$, e.g. using a bulk-boundary
propagator, to act with the bulk generators (\ref{jj2}) on the
exact solutions and to extract the new asymptotic behavior
$(\phi')^{<,>}(\xi,\theta)$. Luckily for the generators
(\ref{jj2}) we can work with the asymptotic behavior only and we
find that
\begin{eqnarray}
M_{ij} & = & \xi_i\frac{\partial}{\partial \xi_j} - \xi_j
 \frac{\partial}{\partial \xi_i} \nonumber \\
K^<_i & = & \frac{\xi_i}{\xi} \left(
\xi\frac{\partial}{\partial\xi}\right) +
  \left( \xi \frac{\partial}{\partial \xi_i} - \frac{\xi_i}{\xi}
 \sum_k \xi_k \frac{\partial}{\partial \xi_k}\right) \nonumber \\
K^>_i & = &\frac{\xi_i}{\xi}
\left(-d-\xi\frac{\partial}{\partial\xi}\right) +
  \left( \xi \frac{\partial}{\partial \xi_i} - \frac{\xi_i}{\xi}
 \sum_k \xi_k \frac{\partial}{\partial \xi_k}\right) \label{jj3}
 \end{eqnarray}
where as we did above we combined the angles and $\xi$ into a
$d+1$ component vector $\xi^i$ with $\xi^2 = \sum \xi_i^2$, and
$K^<_i$ describes the action of the boost generators on
$\phi^<(\xi,\theta)$, and similarly for $K^>_i$ and $\phi^>(\xi,\theta)$.

The form of the generators (\ref{jj3}) is consistent with the fact
that the operators ${\phi}_{\lambda}(\theta)$ are conformal
operators with conformal weights $\lambda$ or $d-\lambda$. It is
interesting to observe that the generators (\ref{jj3}) are simple
differential operators when acting on $\phi^{<,>}(\xi,\theta)$.
Therefore, they give rise to Ward identities for correlation
functions in the dual theory. For example, a two point function of
the form
\be \langle {\phi}^<(\xi_1,\theta_1) {\phi}^<(\xi_2,\theta_2)
\rangle \label{jj4}
\ee
should be annihilated by $K^<_{i,1} + K^<_{i,2}$. In conformal
field theory the conformal Ward identities fix the form of a
two-point function up to normalization, but here it only fixes it
up to one unknown function. Indeed, one can show that the Ward
identities imply that (\ref{jj4}), up to contact terms, is of the
form
\be
\langle {\phi}^<(\xi_1,\theta_1) {\phi}^<(\xi_2,\theta_2)
\rangle = f_<(\xi_1 \xi_2 - \sum_i \xi^i_1 \xi^i_2) \equiv f_<(\xi_1
\xi_2 (1-\cos \gamma(\theta_1,\theta_2))) \label{jj5}
\ee
with some unknown function $f_<$. In terms of the operators 
${\phi}_{\lambda}(\theta)$ the corresponding statement is that the
two-point function is proportional to
$\delta(\lambda_1-\lambda_2)$, but with a $\lambda$-dependent
normalization, so that also in that formulation there is a
function of one variable undetermined.

Similarly, one finds that (again up to contact terms)
\be
\langle {\phi}^<(\xi_1,\theta_1) {\phi}^>(\xi_2,\theta_2)
\rangle = \xi_2^{-d} f(\xi_1 \xi^{-1}_2 (1-\cos
\gamma(\theta_1,\theta_2)))  \label{jj6}
\ee
and
\be
\langle {\phi}^>(\xi_1,\theta_1) {\phi}^>(\xi_2,\theta_2)
\rangle = (\xi_1\xi_2)^{-d} f_>(\xi_1^{-1} \xi^{-1}_2 (1-\cos
\gamma(\theta_1,\theta_2))) ~~. \label{jj61}
\ee
To determine the form of the functions $f$, $f_>$ and $f_<$  
one needs further input.
A typical form of a two-point function is obtained by for example
taking the first term proportional to $A(\lambda)$ in (\ref{p8}),
and by dropping the $e^{-\lambda(\tau+\tau')}$ that appears there.
The integral over $\lambda$ can be done explicitly, using the fact
that the inverse Mellin transform of $\Gamma(a+z) \Gamma(b-z)$ is
$\Gamma(a+b) x^a (1+x)^{-a-b}$ \cite{mellin}. This results
\be
\langle {\cal O}^<(\xi_1,\theta_1) {\cal O}^>(\xi_2,\theta_2)
\rangle = \xi_2^{-d} \left( 1 + \frac{\xi_1}{\xi_2} (1-\cos
\gamma(\theta_1,\theta_2))\right)^{-d/2}
\ee
which is very similar to the asymptotic form of the Feynman
propagator in coordinate space as $\tau,\tau'\rightarrow \infty$.
Similarly, we find  that
\be
\langle {\cal O}^>(\xi_1,\theta_1) {\cal O}^>(\xi_2,\theta_2)
\rangle = (\xi_1\xi_2)^{-d} {(\xi_1{\xi_2})^{d/2} \over (1-\cos
\gamma(\theta_1,\theta_2))^{d/2}}~~.
\lb{jjj}
\ee
Note, that calculating the correlation function (\ref{jjj}) with the 
help of the inverse Mellin transform of product of two Gamma functions
the result would contain the product of the delta function 
$\delta^{(d)}(\theta_1, \theta_2)$ and $\Gamma (0)$. 
This product, however,
with the help of (\ref{A2}) can be recognized as  the kernel
$(1-\cos\gamma(\theta_1,\theta_2))^{-d/2}$.

The Lorentz or conformal generators (\ref{jj2}) took  a very
simple form (\ref{jj3}) when expressed in the new coordinate
$\xi$. A crucial ingredient in the derivation of (\ref{jj3}) was
the fact that we were considering a massless scalar field. In
general we would expect the conformal multiplet and the action of
the conformal generators to depend on the field under
consideration. For example, we can consider a massive scalar field
with its decomposition given in terms of MacDonald functions in
(\ref{mass}). Unfortunately, there does not seem to be a simple
first order differential operator with eigenvalue $\nu$ when
acting on $K_{\nu}(M \xi)$. Therefore, the massive generalization
of (\ref{jj3}) will be very complicated. The mode functions
$\phi_m(x)$ in (\ref{mass}) are still solutions of the massive
field equations on the AdS/dS slices, and the conformal group
still acts in a simple way on these modes. We therefore believe
that ultimately the description in terms of an infinite number of
conformal fields is more useful than the description in terms of
the additional coordinate $\xi$.

Besides Lorentz invariance, the bulk theory is also
translationally invariant. Since our construction relies on a
choice of light-cone, which is not invariant under translations,
we may expect that translational invariance is not very manifest
in our framework. Perhaps this is what is to be expected for a
holographic dual description of Minkowski space, which is
inherently non-local and employs one dimension less.

The explicit form of the translation generators, written in the
same coordinates as used in (\ref{jj2}), reads
\begin{eqnarray}
P^0 & = & \frac{\cosh \tau}{r} \frac{\partial}{\partial \tau} -
 \frac{\sinh \tau}{r} \sum_k r_k \frac{\partial}{\partial r_k}
 \nonumber \\
P^i & =& -\frac{r_i}{r} \frac{\sinh \tau}{r}
\frac{\partial}{\partial \tau} +\frac{1}{r \cosh \tau} \left( r
\frac{\partial}{\partial r_i} \right) + \frac{r_i}{r^2}
\frac{\sinh^2 \tau}{\cosh \tau} \sum_k r_k
\frac{\partial}{\partial r_k} .
\end{eqnarray}
Consider now the first term in (\ref{jj1}), i.e. the one involving
${\cal O}^<(\theta)$. Acting with $P^0$ on this we obtain
\be
{1\over 2\pi i}\int_{{d\over 2}-i\infty}^{{d\over 2}+i\infty}
d\lambda N_{\lambda}^{-1} (-\lambda) r^{-\lambda-1}
e^{-(\lambda+1)\tau} ~ {_{\tt out}}\!{\cal O}^<_\lambda (\theta)
\label{jj8}.
\ee
If we would drop the $e^{-(\lambda+1) \tau}$ exponentials, we would
from this deduce that
\be P^0 \phi^<(\xi,\theta) = \frac{\partial}{\partial \xi}
\phi^<(\xi,\theta) \label{moma}.
\ee
However, this answer cannot be correct. For example, any two-point
function of the form (\ref{jj5}) cannot be annihilated by
$P^0_1+P^0_2$ unless it vanishes identically. The mistake we made
is to work only with the asymptotic form of the fields, neglecting
subleading pieces that are necessarily there in the exact solution
to the field equation. Subleading terms can mix with the result in
(\ref{jj8}), and this changes the action of $P^0$. We have not
worked out the full detailed form of $P^0$ and $P^i$, but a useful
perspective on these operators is obtained by looking at plane
waves in the decomposition given in (\ref{Mellin}) and in
particular in (\ref{p19}). Given a plane wave with momentum $k$,
we denote the associated $\xi,\theta$-dependent fields by
$W^{<,>}_k(\xi,\theta)$. One can work out the explicit form of
these fields. One finds that $W^>_k(\xi,\theta)$ is some function
of $k_0\xi$ times $\delta^{(d)}(\theta_k,\theta)$, where as before
$\theta_k$ indicates the point on the $d$-sphere given by the
$d+1$ dimensional unit vector $k_i/k_0$ (recall that we are still
discussing massless scalar fields here). On the other hand, we
have that
\be W^<_k(\xi,\theta) = \exp(\frac{i}{2}(k_0 \xi - \sum_i k_i
\xi_i)) \label{jj10} .
\ee
This shows that $P^0$ cannot be equal to $\partial/\partial \xi$.
A plane wave has $P^0$ eigenvalue $ik_0$, whereas
$\partial/\partial\xi$ acting on (\ref{jj10}) has eigenvalue
\be
\frac{\partial}{\partial \xi} W^<_k(\xi,\theta) = \frac{i}{2\xi}
(k_0 \xi - \sum_i k_i \xi_i) .
\ee
There is no simple operator for which (\ref{jj10}) has eigenvalue
$ik_0$. Therefore, the easiest way to study translation invariance
would be to rewrite the fields $\phi^<(\xi,\theta)$ in terms of
momentum eigenstates such as (\ref{jj10}), in terms of which
translation invariance simply boils down to momentum conservation.
Incidentally, the fields (\ref{jj10}) are, up to a factor of two,
simply the restriction of a plane wave to the light-cone. This
suggest yet another interpretation for $\xi$, namely as a null
coordinate along the light-cone. Since all AdS and dS slices
asymptote to the light cone, this is perhaps the most natural
interpretation of $\xi$.

A further input on the possible structure of the translations
can be gained from the looking at the explicit form of the conformal 
operators (\ref{p16}) given in section 5.2. 
We find that
\be
&&\phi^>(\xi,\theta)=\int{d\lambda\over 2\pi i}\xi^{-\lambda}\phi^{d-\lambda}(\theta)
\nonumber \\
&&=\xi^{-d/2}\int_0^\infty dk k^{d-1/2}(i^{-d/2}e^{-ik\xi}a_k(k,\theta)+i^{d/2}e^{ik\xi}a^+_k(k,\theta))~~,
\lb{jjh}
\ee
where in the second line we take explicitly the integral over $\lambda$.
Under the time translation $X_0\rightarrow X_0+b$ 
the creation/annihilation operators (\ref{p11})
transform as $a(k,\theta_k)\rightarrow e^{-ikb}a(k,\theta_k)$,
$a^+(k,\theta_k)\rightarrow e^{+ikb}a^+(k,\theta_k)$. 
This induces a transformation for the operators $\phi^>(\xi,\theta)$,
\be
\phi^>(\xi,\theta)\rightarrow {\phi^>}'(\xi,\theta)=(\xi+b)^{d/2} \xi^{-d/2}\phi^>(\xi+b,\theta)~~.
\lb{jjhg}
\ee
The two-point function of the operators (\ref{jjh}) is given by
(\ref{jjj}) and is obviously invariant under (\ref{jjhg}).
A similar analysis for the $<$-operators gives rise to the
form
\be
\phi^<(\xi,\theta)=\int_0^\infty dk k^{d-1/2}
\int d\mu(\theta_k)(e^{-i{k\xi\over 2}(1-\cos\gamma(\theta,\theta_k))}a_k(k,\theta_k)+e^{i{k\xi\over 2}
(1-\cos\gamma(\theta,\theta_k))} a^+_k(k,\theta_k))~~.
\lb{jjgg}
\ee
However the explicit way of how translations act  on the operators
(\ref{jjgg}) is not transparent.
The problem of finding this transformation eventually boils down to the
above mentioned problem of finding the operator for which the exponent
(\ref{jj10})
would have an eigenvalue  $ik$.

Besides translations, there are infinitely many other operators in
the BMS group which are probably also very complicated; naively
they seem to act on fields $\phi^<(\xi,\theta)$ as
$f(\theta)\partial_\xi$, but as the example of $P^0$
demonstrated, this is not quite correct. In addition the BMS
generators are not exact symmetries of Minkowski space, and we
therefore do not expect them to annihilate the vacuum state of the
dual theory, and for the same reason we also do not expect them to
give rise to new Ward identities. At best they provide a set of
spectrum generating operators, but more work is needed to verify
whether this is indeed the case or not. Actually, this applies to
all asymptotic symmetries described in the beginning of this
section, not just to the BMS operators.

\section{Gravitational holographic description}
\setcounter{equation}0
\subsection{General remarks}
In section~4 of this paper we have seen how for a scalar field in
Minkowski space a dual description emerges by first reducing the
scalar field to a set of massive fields on a the constant
curvature (de Sitter or anti-de Sitter) slices of space-time. In
this section we will attempt to make a first step towards
generalizing this duality to the gravitational field. The field
equations in this case are the usual non-linear Einstein equations
(with vanishing cosmological constant)
\be
{\cal R}_{\alpha\beta}=0~~,
\lb{7.1}
\ee
which define a Ricci-flat space-time, which appears to be the
appropriate generalization of flat Minkowski space. Equation
(\ref{7.1}) should be solved subject to suitable boundary
conditions for the metric. We choose so-called normal coordinates
in which the metric has the form
\be
ds^2=dr^2+r^2g_{ij}(x,r) dx^idx^j~~.
\lb{7.2}
\ee
Notice that for concreteness we consider the region analogous to
the region ${\cal D}$ of Minkowski space, where the coordinate $r$
plays the role of a radial coordinate and the metric $g_{ij}(x,r)$
is assumed to have an asymptotic expansion in powers of $1/r$. The
coordinates $\{x^i\}, \ i=1,..,d+1$ are the coordinates on the
slices, in the previous section  we had $x^i=\{\tau,\theta\}$.

The bulk diffeomorphisms preserving the form (\ref{7.2}) are
generated by a vector field $\xi=(\xi^r,\ \xi^i)$ such that
\be
&&\xi^r=\alpha (x)~,~~\xi^i=\xi^i(x,r) \nonumber \\
&&\partial_r\xi^i=-{1\over r^2}g^{ij}(x,r)\partial_j\alpha (x)~~.
\lb{df}
\ee
The metric components change according to
\be
{\cal L}_\xi g_{ij}(x,r)=\alpha (x)r^{-2}\partial_r(r^2g_{ij})+\nabla^{(g)}_i\xi_j+
\nabla^{(g)}_j\xi_i ~~.
\lb{df1}
\ee

The first term  $g^{(0)}_{ij}(x)$ in the expansion of
$g_{ij}(x,r)$ is the metric on the asymptotic space-like boundary
of space-time. In analogy with the AdS/CFT correspondence one
could try to keep the metric $g^{(0)}_{ij}(x)$ fixed and one could
try to solve the bulk Einstein equations subject to this Dirichlet
condition. However, in the present case when the bulk metric is
determined by equation (\ref{7.1}) the metric $g^{(0)}_{ij}(x)$ is
not arbitrary but has to satisfy (see \cite{SS}, \cite{BS},
\cite{B})
\be
R_{ij}[g^{(0)}]=d~ g^{(0)}_{ij}~~.
\lb{7.3}
\ee
This means that the boundary at space-like infinity is described
by an Einstein metric of constant positive curvature.

Provided that a metric $g^{(0)}_{ij}(x)$ satisfying (\ref{7.3}) is
fixed one can try to develop the $1/r$ expansion of the metric
(\ref{7.2}) and determine the other terms in the expansion of
$g_{ij}(x,r)$. To some extent this was analyzed in
\cite{deHaro:2000wj} and it was found that the relation between
the coefficients in the expansion of $g_{ij}(x,r)$ is differential. 
This is different from what happens in asymptotically
Anti-de Sitter space, where the relations are algebraic \cite{HS}. This
makes it difficult to solve the recurrence relations and to
determine the coefficients in the series provided the first
coefficient $g^{(0)}_{ij}(x)$ is given.

However, the holographic point of view we are trying to develop in
this paper is not based on fixing boundary data at spatial
infinity. It is based on fixing an infinite set of data near the
boundaries of the light-cone, and the holographic reduction takes
place along the anti-de Sitter slices, and not along the
coordinate $r$. Thus, what we should really try to is to take the
metric in the form (\ref{6.1}), and to study the field equations
for this metric in this form. Ideally, there should exist some
separation of variables that allows us to separate the
$r$-dependence from the dependence on the other coordinates, just
as we did for the massless scalar field. In view of the
non-linearity of the Einstein equations it is not clear exactly
how we should separate the $r$-dependence, but assuming we have
achieved this the $r$-dependence will be gone and it will be
replaced by a dependence on some spectral parameter $\lambda$. The
boundary data will then involve quantities such as
$\sigma^{\lambda,\pm}(\theta)$, $g^{\lambda,\pm}_{ij}(\theta)$ and
$A^{\lambda,\pm}_i(\theta)$. The holographic reconstruction of the
metric now requires us to determine the $\rho$-dependence, and
those equations are indeed algebraic. We leave a more detailed
study of these issues to future work and limit ourselves here to
the discussion of one example, namely the relatively simple case
of $(2+1)$-dimensional Minkowski space.

\subsection{Example: (2+1)-dimensional Minkowski space-time}

In three space-time dimensions the Riemann tensor is algebraically
related to the Ricci tensor. This property was used in \cite{KS}
to find a general solution to the three-dimensional Einstein
equations with a negative cosmological constant.

In the case at hand, the fact that the Ricci tensor vanishes
implies that the Riemann tensor vanishes as well,
\be
{\cal R}_{\alpha\beta\mu\nu}=0~~,
\lb{7.4}
\ee
i.e. space-time is locally flat which is of course a well-known
fact in three dimensions. This is a great simplification in
solving (\ref{7.1}) and in fact allows us to find an exact
solution in a way similar to the analysis given in \cite{KS}.

By introducing the variable $\gamma_{ij}(x,r)=r^2g_{ij}(x,r)$ the
equation (\ref{7.4}) for the metric (\ref{7.2}) reduces to the
following set of equations
\be
\gamma''={1\over 2}\gamma'\gamma^{-1}\gamma'~~,
\lb{7.5}
\ee
\be
R_{likj}[\gamma]={1\over 4}\gamma'_{ij}\gamma'_{lk}-{1\over 4}\gamma'_{ik}
\gamma'_{lj}~~,
\lb{7.6}
\ee
\be
\nabla_k^{(\gamma)}\gamma'_{ij}-\nabla_j^{(\gamma)}\gamma'_{ik}=0~~,
\lb{7.7}
\ee
where we denote $\gamma'=\partial_r\gamma$.
Differentiating equation (\ref{7.5}) once again with respect to $r$
we find that
$$
\gamma'''=0~~.
$$
Hence $\gamma$ is a quadratic function of $r$, i.e.
\be
\gamma(x,r)=r^2g_{(0)}+rg_{(1)}+g_{(2)}~~.
\lb{7.8}
\ee
Inserting this back into equation (\ref{7.5}) we find that
\be
g_{(2)}={1\over 4}g_{(1)}g^{-1}_{(0)}g_{(1)}
\lb{7.9}
\ee
so that (\ref{7.8}) can be written as a total square
$$
\gamma (x,r)=r^2\left(1+{1\over 2r}g_{(1)}g^{-1}_{(0)}\right) g_{(0)}
\left(1+{1\over 2r}g^{-1}_{(0)}g_{(1)}\right)~~.
$$
Next, to solve equation (\ref{7.6}) we use the following identity,
valid in two dimensions, that relates the Riemann tensor to the
Ricci scalar,
$$
R_{likj}[\gamma]={1\over 2} R[\gamma]
\left(\gamma_{kl}\gamma_{ij}-\gamma_{jl}\gamma_{ki}\right)~~.
$$
Inserting (\ref{7.8}) into eq.(\ref{7.6}), expanding in powers of $1/r$
and using the above identity we find  in the leading order
that
\be
R[g_{(0)}]=2~~,
\lb{7.10}
\ee
i.e. the space at infinity is two-dimensional de Sitter space.
This is a particular manifestation of the general behavior
(\ref{7.3}) of the asymptotically flat metric. At the next order
in $1/r$ equation (\ref{7.6}) reduces to a single
equation\footnote{Hereafter we use $g_{(0)}$ to define the trace,
e.g. $\Tr g_{(1)}=\Tr(g^{-1}_{(0)}g_{(1)})$.}
\be
\nabla^i_{(0)}\nabla^j_{(0)}g_{(1)ij}-\nabla^2_{(0)} \Tr g_{(1)}=0~~.
\lb{7.11}
\ee
On the other hand, equation (\ref{7.7}) is trivially satisfied to
the leading order and in the next order gives rise to the equation
\be
\nabla^k_{(0)}g_{(1)ki}=\partial_i \Tr g_{(1)}~~.
\lb{7.12}
\ee
We see that (\ref{7.11}) is  a  consequence of (\ref{7.12}).
The equations (\ref{7.10}) and (\ref{7.12}) are the only
constraints to be imposed on the terms in the expansion (\ref{7.8}).

Notice that equation (\ref{7.12}) means that the tensor
$g_{(1)ij}-g_{(0)ij} \Tr g_{(1)}$, like a stress-energy tensor, is
covariantly conserved with respect to the metric $g_{(0)ij}$. 
In fact, according to the prescription of Brown
and York \cite{BrownYork} the quasi-local stress-energy tensor of
a spacetime is defined on the boundary (placed at a fixed value of
$r$ which then should be taken to infinity) as a variation of the
gravitational action with respect to the boundary metric and takes
the form
\be
T^{\tt bound}_{ij}[\gamma]&=&-{1\over 8\pi G}(K_{ij}-K\gamma_{ij}) \nonumber \\
&=&-{1\over 8\pi G}\left(-2r g_{ij}+
r^2(\partial_r g_{ij}-g_{ij}\Tr(g^{-1}\partial_r g)\right)~~.
\lb{7.13}
\ee
This expression is divergent when $r$ is taken to infinity. This
can be thought of as a UV divergence in the theory on de Sitter space. It
can be regularized by adding a local counterterm
\be
T^{\tt ct}_{ij}[\gamma]={1\over 4\pi G}{1\over r} \gamma_{ij}~~.
\lb{7.14}
\ee
The total stress tensor defined as the sum of the two
contributions,
\be
T_{ij}[\gamma]&=&T^{\tt bound}_{ij}+T^{\tt ct}_{ij} \nonumber \\
&=& -{1\over 8\pi G}(g_{(1)ij}-g_{(0)ij} \Tr g_{(1)})
\lb{7.15}
\ee
then has a finite limit when $r$ is taken to infinity and this
limit reproduces the expression in the second line of
(\ref{7.15}). We see that up to a factor the stress tensor
$T_{ij}$ is exactly the tensor which enters (\ref{7.12}) and thus
should be covariantly conserved with respect to the metric
$g^{(0)}_{ij}$,
\be
\nabla^j_{(0)}T_{ij}=0~~.
\lb{T}
\ee
In the remainder of the discussion we will use units in which
$8\pi G=1$.

Although empty $(2+1)$-space-time is locally flat it becomes
curved in the presence of matter. For instance, a point particle
creates $\delta$-like conical singularity distributed along the
particle world line. The stress-energy tensor (\ref{7.15}) then
determines the energy and angular momentum of the matter as
measured at space-like infinity. In terms of $T_{ij}$ we have that
\be
g^{(1)}_{ij}=-T_{ij}+g^{(0)}_{ij}\Tr T~~.
\lb{g1}
\ee
Taking into account (\ref{7.8}) and (\ref{7.9}) we see that the
boundary stress tensor $T_{ij}(x)$ and the boundary metric
$g^{(0)}_{ij}(x)$ subject to constraint (\ref{7.10}) completely
determine the bulk metric satisfying the Ricci-flat equation
(\ref{7.1}) in three dimensions. This is  similar to what we had
in the case of negative cosmological constant in three dimensions
\cite{KS}. The difference is that in the latter case the trace of
the boundary stress tensor is fixed and no constraint on the
boundary metric arises. The stress tensor on the de Sitter
boundary of asymptotically Minkowski space-time  should be
covariantly conserved with respect to the metric $g^{(0)}_{ij}$
but is otherwise arbitrary. In particular, there is no restriction
on the trace of $T_{ij}$. This would be quite natural if we were
doing a holographic reduction in the radial direction, since then
$T_{ij}$ should represent a stress tensor of a boundary theory
which is holographically dual to a theory in the bulk. But we
emphasize once more that this not the type of holographic
reduction we are attempting to do in this paper.

To the leading order the $r$-dependent diffeomorphism (\ref{df})
is generated by the vector field
\be
\xi^r=\alpha(x)~~,~~~\xi^i={1\over r}g^{ij}_{(0)}\partial_j\alpha (x)~~,
\lb{xir}
\ee
where $\alpha(x)$ is arbitrary function on the de Sitter boundary.
The boundary metric $g^{(0)}_{ij}$ remains unchanged under this transformation while
$g^{(1)}_{ij}$ transforms  as
\be
{\cal L}_\xi g^{(1)}_{ij}=\alpha (x) g^{(0)}_{ij}+\nabla^{(0)}_i
\nabla^{(0)}_j\alpha~~.
\lb{7.16}
\ee
This transformation
acts as
\be
\delta_\alpha T_{ij}=-\nabla^{(0)}_i
\nabla^{(0)}_j\alpha +g^{(0)}_{ij}\nabla_{(0)}^2\alpha +\alpha g^{(0)}_{ij}
\lb{7.17}
\ee
on the stress tensor $T_{ij}$. It is easy to check that the r.h.s.
in (\ref{7.17}) is covariantly conserved on a two-dimensional
space-time of constant curvature (\ref{7.10}). It can be obtained
by varying the following ``dilaton-gravity'' boundary action
\be
W_{\tt \alpha}=\int d^2x\sqrt{g_{(0)}}~\alpha (x) ~( R[g_{(0)}]-2)~~
\lb{7.18}
\ee
with respect to the metric. Note also that the variation of
(\ref{7.18}) with respect to $\alpha(x)$ gives rise to the
equation (\ref{7.10}).

Let us now show that on de Sitter space the stress tensor $T_{ij}$
can be  determined by solving (\ref{T}) provided its trace is some
known function
\be
\Tr T=C(x)~~.
\lb{trT}
\ee
We choose the metric
\be
ds^2={1\over z^2}(-dz^2+d\theta^2)={4\over (x_+-x_-)^2}dx_+dx_-~~,
\lb{dS}
\ee
where $x_+=\theta+z$, $x_-=\theta-z$,  to describe the
two-dimensional de Sitter space. This coordinate system covers
half of de Sitter space, the boundary then being at $z=0$. The
trace equation (\ref{trT}) then determines the $(+-)$ component of
the stress tensor
\be
T_{+-}={1\over (x_+-x_-)^2}C(x_+,x_-)~~.
\lb{7.19}
\ee
From (\ref{T}) we obtain a couple of differential equations
\be
\partial_-T_{++}=-{1\over (x_+-x_-)^2}\partial_+C ~,~~
\partial_+T_{--}=-{1\over (x_+-x_-)^2}\partial_+C~~.
\lb{7.20}
\ee
A solution can be  easily found  and it takes the form
\be
T_{++}(x_+,x_-)=f(x_+)-\int^{x_-}{dv\over (x_+-v)^2}\partial_+ C(x_+,v)
\nonumber \\
T_{--}(x_+,x_-)=h(x_-)-\int^{x_+}{du\over (x_--u)^2}\partial_- C(u,x_-)~~,
\lb{7.21}
\ee
where $f(x_-)$ and $h(x_+)$ are arbitrary functions of $x_-$ and
$x_+$ respectively. The boundary stress tensor (\ref{7.19}),
(\ref{7.21}) and eventually the bulk metric are determined by
specifying the  functions $f(x_+)$, $h(x_-)$ and $C(x_+,x_-)$. Let
us now return to the holographic interpretation of this solution.
The central idea proposed in this paper is to reduce everything to
holographic data on the boundary of the light-cone, which in the
present case is a circle located at $z=0$ and parametrized by the
angle coordinate $\theta$. Thus, from the point of view of the
holographic reconstruction it is important to see what the minimal
data is that we need to fix on this circle in order to reconstruct
the full $(2+1)$-dimensional spacetime. We see that the values of
the functions $f(x_+)$ and $h(x_-)$ at late times $z$ are
completely determined by their values at $z=0$, i.e. $f(\theta)$
and $h(\theta)$. On the other hand, the function $C(x_+,x_-)$
seems to require an infinite set of functions to be specified at
$z=0$, namely all coefficients in the expansion
\be
C(x_+,x_-)=\sum_{n=0}^\infty C_n(\theta) z^n~~,
\lb{Cn}
\ee
where $\theta={1\over 2}(x_++x_-)$ and $z={1\over 2}(x_+-x_-)$.
However, most of this infinite set of functions can be gauged away
by the coordinate transformation (\ref{xir}). As is seen from
(\ref{7.17}) this transformation parametrised by
$$
\alpha(x_+,x_-)=\sum_{n=0}^\infty \alpha_n(\theta)z^n
$$
transforms the function $C(x_+,x_-)$. Specifically, the coefficients in the expansion
(\ref{Cn}) transform as follows
\be
\delta_\alpha C_n(\theta )=
{1\over 4}\alpha''_{n-2}(\theta)-(n-2)(n+1)\alpha_n(\theta)~~.
\lb{dCn}
\ee
We see that the parameter function $\alpha(x_+,x_-)$ contains
enough freedom to eliminate almost all functions $C_n(\theta)$
with one exception: the same parameter $\alpha_0(\theta)$ is used
to transform both $C_0(\theta)$ and $C_2(\theta)$ so that one can
not eliminate both. In fact the combination
$$
c(\theta)=C_2(\theta)-2C''_0(\theta)
$$
is invariant under the transformation, $\delta_\alpha
c(\theta)=0$. Thus, the functions $c(\theta)$ together with
$f(\theta)$ and $h(\theta)$ form the only independent and
``gauge-invariant'' set of data which should be specified at the
boundary of the light-cone. These data are sufficient to
holographically reconstruct the whole three-dimensional metric.
Actually, this is a quite subtle statement. Asymptotic
diffeomorphisms do not in general extend to smooth diffeomorphisms
of the entire space, and acting with them typically generates
singularities. The diffeomorphisms used to reduce the infinite set
of functions $C_n(\theta)$ to just one function of $\theta$ are
presumably of this form. Therefore, if we are interested in smooth
solutions, it is probably better to work with the infinite set
$C_n(\theta)$ instead. For this infinite set the existence of a
globally smooth solution is obviously also not guaranteed, and
more work is required to determine which sets of $C_n(\theta)$
correspond to smooth solutions.

We finish this section with a comment on a RG interpretation of
the $(2+1)$-dimensional metric. As we have seen above, if we view
$r$ as the direction in which we do the holographic reduction the
standard interpretation of the radial position $r$ of the boundary
is as a UV cut-off in the boundary theory. The dependence of the
metric on $r$ can be viewed as some kind of RG-flow. In this
respect it is interesting to see whether this flow is irreversible
so that some kind of analog of the C-theorem could be found.
Remarkably the flow, as it is dictated by the bulk Einstein
equation (\ref{7.1}), is indeed irreversible. In order to see
this, we consider the quasi-local stress tensor (\ref{7.13}) and
examine how its trace computed with respect to the induced metric
$\gamma_{ij}$ changes as a function of $r$. Using equation
(\ref{7.5}) we find that
\be
\partial_r \Tr T^{\tt bound}=-{1\over 2}\Tr
(\gamma^{-1}\gamma'\gamma^{-1}\gamma')<0~~.
\lb{7.22}
\ee
However, it is not clear whether the bulk gravitational dynamics
drives the flow to a fixed point which would correspond to a
conformal field theory on de Sitter boundary of Minkowski space.
In any case this is not the point of view we want to pursue
further.

\section{Conclusion}
\setcounter{equation}0

In this paper we have proposed a holographic reduction of
Minkowski space. The central idea was to perform a holographic
reduction along the Euclidean anti-de Sitter and de Sitter slices
in which Minkowski space is naturally foliated. The end product is
that we obtained an infinite set of conformal operators on the two
spheres that form the conformal boundary of the light-cone. For a
massless scalar field in the bulk we gave a precise definition of
a Minkoswki/CFT duality which also extends to interacting fields.
We have mainly considered cases where we took all points either
inside Euclidean anti-de Sitter space, or inside de Sitter space,
but the results in this paper can be extended to the mixed case as
well.

It is perhaps worth pausing for a moment to see what we have
learnt about de Sitter space holography from this approach. The
first lesson seems to be that the natural conformal weights to
emerge on de Sitter space are complex ones. Since a massless
scalar field in Minkowski space describes a unitary theory, and
carries a unitary representation of the Lorentz group, the
corresponding theory on de Sitter space and its holographic dual
should also be unitary. All this is evidence in favor of the
picture of the dual of de Sitter space advocated in
\cite{Balasubramanian:2002zh}. There, it was proposed that the
dual CFT has non-standard hermiticity conditions, such that
complex conformal weights do not violate unitarity, and
$SO(d+1,1)$ is unitarily realized. In fact, there are two CFT's,
one living on each boundary, and full correlation functions are
given by evaluating them in the product theory in a suitable
entangled state. This entangled state is the analogue of the
S-matrix, and the set of correlation functions on each of the
spheres at infinity represents the in and out Hilbert spaces.

Clearly, it should also be possible to study other manifolds that
can be sliced in (anti)-de Sitter slices, and to study holography
in those cases. It would also be interesting to study e.g. anti-de
Sitter space in the Poincare patch, to view this as a slicing with
Minkowski slices and to see if the known holography for anti-de
Sitter space is consistent with our proposed Minkowski space
holography.

A further extension and a crucial test of our idea would be to
study the Schwarzschild metric, and to see how the entropy of the
Schwarzschild black hole is reflected in the dual CFT. For this,
we should first further develop the extension of our method to the
gravitational field, for which only a few first steps were made in
sections~6,7. Especially, it is not yet clear exactly which data
at infinity we should specify in order to reconstruct the bulk
geometry, and whether this data allows for an interpretation in
terms of an infinite set of stress-energy tensors.

One can imagine many other extensions. We have only briefly
considered massive scalar fields, and it would be desirably to
understand these better. Another thing that comes to mind is a
supersymmetric generalization (or a generalization to string
theory). Since supersymmetries close into translation generators
and we already saw that translation symmetry does not take on a
very simply form, we suspect that it will also be difficult to
keep supersymmetry manifest in the formulation. The fact that
translation invariance is implemented in such a cumbersome way is
perhaps to be expected for a holographic dual description of
Minkowski space, but it would be nice to have a clearer
understanding of this point.

Even though free massive particles will not approach the boundary
of the light-cone, but rather go to timelike infinity, we still
saw that it is possible to describe them in terms of data on the
boundary of the light cone. The situation gets even more confusing
for strongly interacting degrees of freedom, such as those that
appear in confining gauge theories, and for which the LSZ
formalism does not seem appropriate. We nevertheless still expect
that the physics of such degrees of freedom will be captured by
suitable data on the two spheres that form the boundary of the
light-cone.

Of course, we are not living in Minkowski space, and it is an even
more urgent question to find a holographic dual description of a
realistic time-dependent cosmological solution. Nevertheless,
Minkowski space is still a good approximation for many physical
questions, and understanding Minkowski space is definitely a step
in the right direction. Incidentally, four-dimensional Minkowski
space reduces in this philosophy to a theory on two two-spheres,
which have an infinite dimensional conformal symmetry, and this
may lead to an interesting role of the Virasoro algebra for
four-dimensional physics.

\vspace{1cm} \noindent {\bf ACKNOWLEDGEMENTS} We would like to
thank V.~Balasubramanian, C.~Hull, N.~Lambert, D.~Minic, H.~Reall,
K.~Skenderis, M.~Taylor-Robinson and E.~Verlinde for discussions.
S.S. would like to thank ITFA for hospitality during final stage 
of this work. Research of S.S. is supported by the grant 
DFG-SPP 1096, Stringtheorie.

\appendix{Kernels and delta-functions}
\label{kernels}
\setcounter{equation}{0}
The kernel
$$
(\cosh y-\sinh y\cos\gamma(\theta,\theta'))^{-\lambda}
$$
produces a delta-function concentrated at $\theta=\theta'$ when
$y\rightarrow \infty$. In order to see this and get the precise
factor we take $\theta\simeq \theta'$ and approximate $\cos \gamma
(\theta,\theta')\simeq 1-\gamma^2/2$. In the limit of large $y$
the kernel takes the form
$$
{2^{2\lambda}e^{-\lambda y}\over (\gamma^2+4e^{-2y})^\lambda}~~.
$$
By introducing the small parameter $\delta=2e^{-y}$ this can be
brought in the form
$$
2^d~ e^{-(d-\lambda)y}\left({\delta^{2\lambda-d}\over (\gamma^2(\theta,\theta')+\delta^2)^\lambda}
\right)~~,
$$
where the term in the parenthesis generates the delta-function
$\delta^{(d)}(\theta,\theta')$ on d-sphere up to some
normalization constant. In order to determine this constant we
consider the integral
$$
\lim_{\delta\to 0}\int_{S_d}{\delta^{2\lambda-d}\over
(\gamma^2(\theta,\theta')+\delta^2)^\lambda}
={2\pi^{d\over 2}\over \Gamma({d\over 2})}\int_0^\infty{dx~ x^{d-1}\over (x^2+1)^\lambda}~~,
$$
where the first factor is the area of  a (d-1)-sphere and in the
integration over the azimuthal angle we changed the variable
$\gamma$ to $\delta x$. The $x$-integration results in
$$
\int_0^\infty{dx ~x^{d-1}\over (x^2+1)^\lambda}={1\over 2}{\Gamma({d\over 2})\Gamma (\lambda-{d\over 2})\over \Gamma(\lambda)}~~.
$$
Putting things together we find that
\be
\lim_{y\to\infty}(\cosh y-\sinh y\cos\gamma(\theta,\theta'))^{-\lambda}=
(2\pi )^{d\over 2}2^{d\over 2}{\Gamma(\lambda-{d\over 2})\over
\Gamma(\lambda)}~e^{-(d-\lambda)y}\delta^{(d)}(\theta,\theta')~~.
\lb{A1}
\ee
A similar analysis can be done for the kernels
$$
(\sinh\tau -\cosh \tau\cos\gamma (\theta,\theta')+i\epsilon)^{-\lambda}
$$
in the limits $\tau\rightarrow +\infty$ or $\tau\rightarrow
-\infty$. The integration over the azimuthal angle in this case is
$$
\int_0^\infty{dx ~x^{d-1}\over (x^2-1+i\epsilon)^\lambda}=
\int_0^{+i\infty}{dz ~z^{d-1}\over (z^2-1)^\lambda}=
i^{d-2\lambda}\int_0^\infty {dy ~y^{d-1}\over (y^2+1)^\lambda}~~,
$$
where we first deformed the integration contour to the positive
imaginary axis (the possible integral over the quarter-circle at
infinity vanishes provided $Re(2\lambda-d)$ is slightly positive)
and then changed variables $z\rightarrow iy$ that brings us back
to the already known integral given above. Thus, we find that
\be
&&\lim_{\tau\to+\infty}
(\sinh\tau -\cosh\tau \cos\gamma (\theta,\theta')+i\epsilon)^{-\lambda}\nonumber \\
&&=(2\pi )^{d/2}~2^{d\over 2} ~i^{(d-2\lambda)}~{\Gamma(\lambda-{d\over 2})\over
\Gamma(\lambda)} e^{-(d-\lambda)\tau}\delta^{(d)}(\theta,\theta') ~~.
\lb{A2}
\ee
When $\tau\rightarrow -\infty$ the kernel diverges at
$\gamma(\theta,\theta')=\pi$ and hence delta-function that is
obtained in this limit
\be
&&\lim_{\tau\to-\infty}
(\sinh\tau -\cosh\tau \cos\gamma (\theta,\theta')\pm i\epsilon)^{-\lambda}\nonumber \\
&&=(2\pi )^{d/2}~2^{d\over 2} ~(\mp i)^{d}~{\Gamma(\lambda-{d\over 2})\over
\Gamma(\lambda)} e^{(d-\lambda)\tau}\delta^{(d)}(\theta,\pi-\theta')
\lb{A3}
\ee
is concentrated at the anti-podal points on the d-sphere.

\appendix{Plane wave normalization}
\label{normalization}
\setcounter{equation}{0}
The plane waves
\be
f_k(X_0,{\bf{X}})={1\over [(2\pi)^{d+1}2k]^{1/2}}e^{-i(kX_0-\bf{k}\bf{X})}
\lb{B1}
\ee
are delta-function normalized
\be
(f_k,f_p)=\delta^{(d+1)}({\bf{k}},{\bf{p}})=k^{-d}
\delta(k-p)\delta^{(d)}(\theta_k,\theta_p)
\lb{B2}
\ee
with respect to the standard Klein-Gordon scalar product which is
defined as
\be
(f_k,f_p)=i\int_\Sigma d\sigma (f^*_k\partial_n f_p-\partial_n f^*_k f_p)
\lb{B3}
\ee
provided the hypersurface $\Sigma$ (with the normal vector $n$)
is a surface of constant Minkowski time $X_0$.

However, for our purposes we would like to choose a hypersurface
associated with the boundary of the light-cone. More specifically
we define $\Sigma_{\tt out}$ as a hypersurface  consisting of two
components: a surface $\tau=+\infty$ in the region $\cal D$ and a
surface $y=+\infty$ ($t>0$) in the region $\cal A$. The
hypersurface $\Sigma_{\tt out}$ thus surrounds the boundary
$S^+_d$ of the light-cone. Similarly we define $\Sigma_{\tt
in}=\{{\cal D}:\tau=-\infty\}\cup \{{\cal A}:y=+\infty~(t<0)\}$
near the boundary $S^-_d$.

Here we show that same normalization (\ref{B2}) still holds if we
choose $\Sigma_{\tt out}$ (or $\Sigma_{\tt in}$) in the scalar
product (\ref{B3}). We demonstrate this explicitly for the surface
$\Sigma_{\tt out}$. The non-vanishing components of the normal
vector $n$ are $n^\tau=r^{-1}$ (in region $\cal D$) and
$n^y=t^{-1}$ (in region $\cal A$). Hence each component gives
\be
&&\int_{\Sigma_{\tt out}}d\sigma \partial_n
=\int_{\tau\to+\infty}(\cosh\tau )^d
d\mu(\theta) \int_0^\infty dr~r^{d-1}~\partial_\tau \nonumber \\
&&-\int_{y\to+\infty}(\sinh y )^d
d\mu(\theta) \int_0^\infty dt~t^{d-1}~\partial_y~~
\lb{B4}
\ee
in the scalar product (\ref{B3}). As $y\rightarrow +\infty$ the
plane wave asymptotically behaves as
\be
e^{\pm i(kX_0-\bf{k}\bf{X})}={2^d\pi^{d\over 2}\over 2\pi i}\int d\lambda
~(\pm i)^\lambda \Gamma(\lambda-{d\over 2})~k^{-\lambda}~t^{-\lambda}
e^{-(d-\lambda)y}~
\delta^{(d)}(\theta,\theta_k)~~,
\lb{B5}
\ee
where we again assumed that $Re (\lambda-{d\over 2})$ is slightly
positive which allows us to drop the term proportional to
$e^{-\lambda y}$. Similarly in region $\cal D$ the plane wave
asymptotically behaves for $\tau\rightarrow +\infty$ as
\be
e^{\pm i(kX_0-\bf{k}\bf{X})}={2^d\pi^{d\over 2}\over 2\pi i}\int d\lambda
~(\pm i)^{d-\lambda}
\Gamma(\lambda-{d\over 2})~k^{-\lambda}~r^{-\lambda} e^{-(d-\lambda)\tau}~
\delta^{(d)}(\theta,\theta_k)~~.
\lb{B6}
\ee
Inserting these into the definition of the scalar product
(\ref{B3})-(\ref{B4}) and taking into account that
$$
\int_0^\infty dr ~r^{d-\lambda-\lambda'-1}=2\pi i ~\delta(\lambda+\lambda'-d)
$$
we find
\be
&&(f_k,f_p)_{\Sigma_{\tt out}}=-{(2\pi )^{-1}\over \sqrt{k~p}}p^{-d}~\delta^{(d)}(\theta_k,\theta_p)
\nonumber \\
&&{1\over 2\pi i}\int d\lambda \sin {\pi\over 2}(d-2\lambda)~(d-2\lambda)~
\Gamma(\lambda-{d\over 2})~\Gamma({d\over 2}-\lambda)~\left({k\over p}\right)^{-\lambda}
\lb{B7}
\ee
for the scalar product over $\Sigma_{\tt out}$. Notice that the
powers of $i$ coming from the contributions of each component
nicely combine into
$i^{d-2\lambda}-i^{2\lambda-d}=2i~\sin{\pi\over 2}(d-2\lambda)$.
Using that the Gamma function satisfies
$$
\Gamma(\lambda-{d\over 2})\Gamma({d\over 2}-\lambda)
=-{2\pi\over (d-2\lambda)\sin{\pi\over 2}(d-2\lambda)}~~
$$
the $\lambda$-integration in (\ref{B7}) reduces to a simple
integral
$$
{1\over 2\pi i}\int_{{d\over 2}-i\infty}^{{d\over 2}+i\infty}d\lambda ~\left({k\over p}\right)^{-\lambda}=k~\delta(k-p)~~.
$$
Thus, equation (\ref{B7}) is indeed identical to
(\ref{B2}).

Note that this result could have been anticipated. Indeed, suppose
that we add to the hypersurface $\Sigma_{\tt out}$ the surface
$t=+\infty$ in the region $\cal A$. The resulting hypersurface can
be thought of as a deformation of the surface $X_0=+\infty$ for
which the standard normalization (\ref{B2}) holds. Now, the
Klein-Gordon scalar product (\ref{B3}) should not change under any
deformation of the surface $\Sigma$. On the other hand, the
contribution of the surface $t=+\infty$ to the scalar product
vanishes provided $Re(2\lambda-d)$ is small but positive. Hence,
the whole contribution to the scalar product should come entirely
from the hypersurface $\Sigma_{\tt out}$. We however found it
useful to give an explicit calculation of the scalar product over
$\Sigma_{\tt out}$. This serves both as an illustration of our
techniques as well as a consistency check.

\appendix{Diffeomorphisms}
\label{diffeos} \setcounter{equation}{0} Here we perform analysis
of the diffeomorphisms preserving the asymptotic structure of the
metric (\ref{6.1}). Suppose that the diffeomorphism is generated
by a vector field $\xi=(\xi^t,~\xi^\rho,~\xi^i)$. By requiring
that ${\cal L}_\xi G_{tt}=0$ we obtain $\partial_t\xi^t=0$ from
which we find that the $t$-component of the vector $\xi$ does not
depend on $t$ so that
\be
\xi^t=\alpha (\theta, \rho )~~,
\lb{a1}
\ee
where $\alpha (\theta, \rho )$ is arbitrary function of $\theta $ and $\rho$.

The condition ${\cal L}_\xi G_{t\rho}=0$ leads to
\be
\partial_t \xi^\rho={4\rho^2\over (1+{1\over t}\sigma)}
\left(t^{-2}\partial_\rho \alpha-t^{-1}A_i\partial_t\xi^i\right)~~.
\lb{a2}
\ee
and from the condition ${\cal L}_\xi G_{ti}=0$
we find that
\be
\partial_t\xi^i=\rho\left(t^{-2}g^{ji}\partial_i\alpha-t^{-1}g^{ji}A_i\partial_t\xi^\rho
\right)~~.
\lb{a3}
\ee
Multiplying equation (\ref{a3}) by $A_j$ and inserting the result
into equation (\ref{a2}) we find the equations for the
$t$-derivatives of $\xi^\rho$ and $\xi^i$:
\be
\partial_t \xi^\rho={1\over (1+{1\over t}\sigma-{4\rho^3\over t^2}A^2)}{4\rho^2\over t^2}
\left(\partial_\rho \alpha-{\rho\over t}A^i\partial_i\alpha\right)~~.
\lb{a4}
\ee
\be
\partial_t\xi^i=
{\rho\over t^2}g^{ji}\partial_i\alpha -{4\rho^3\over t^3}
{g^{ji}A_i\over (1+{1\over t}\sigma-{4\rho^3\over t^2}A^2)}
(\partial_\rho\alpha-{\rho\over t}A^i\partial_i\alpha)~~,
\lb{a5}
\ee
where we introduced $A^2=A_ig^{ij}A_j$ and all $i$-indexes are
raised with help of the metric $g_{ij}(t,\theta, \rho )$.

Now, let us make a substitution
\be
&&\xi^\rho=\gamma (\theta, \rho )+{1\over t}\beta (t,\theta, \rho) \nonumber \\
&&\xi^i=\xi^i_{(0)}(\theta, \rho )+{1\over t}\xi^i_{(1)}(\theta, \rho )+..
\lb{a6}
\ee
The $t^2$ term in the $G_{\rho\rho}$ component of the metric is
fixed and should not be changed under the diffeomorphism. This
implies $\rho\partial_\rho\gamma (\theta, \rho)-\gamma (\theta,
\rho)=0$ which in turn leads to
\be
\gamma (\theta, \rho)=\gamma (\theta)\rho~~,
\lb{a7}
\ee
where $\gamma (\theta)$ is an arbitrary function. Similarly, the
$t^2$ term in $G_{i\rho}$ of the metric (\ref{6.1}) vanishes.  In
order to preserve this condition the $t$-independent part of
$\xi^i$ should satisfy
\be
\partial\rho \xi^j_{(0)}=-{1\over 4}g^{ji}_{(0)}\partial_j\gamma (\theta )~~.
\lb{a8}
\ee
The diffeomorphism generated by the vector field
$$\xi=\left(0,~\gamma (\theta )\rho,~ \xi^j_{(0)}(\theta, \rho)\right)
$$
is the only $t$-independent diffeomorphism that preserves the form
(\ref{6.1}).

As for the $t$-dependent parts of the functions, $\alpha (\theta,
\rho )$ plays the role of the parameters of the transformation.
Equations (\ref{a4}) and (\ref{a5}) determine $\beta (t,\theta
,\rho)$ and $\xi^i_{(k>0)}(\theta, \rho)$ in terms of $\alpha
(\theta,\rho )$. For the first terms in the $t$-expansion of these
functions we find that
\be
&&\beta_{(0)}(\theta,\rho )=-4\rho^2\partial_\rho \alpha(\theta, \rho )~, ~~\nonumber \\
&&\beta_{(1)}=2\rho^2\left(\sigma_{(0)}\partial_\rho\alpha+\rho A^i_{(0)}\partial_i\alpha
\right)
\lb{a9}
\ee
and
\be
\xi^i_{(1)}=\rho g^{ij}_{(0)}\partial_j\alpha (\theta, \rho )~~.
\lb{a10}
\ee

The subleading terms in $G_{\rho\rho}$ and $G_{i\rho}$ change
under the diffeomorphisms and will give us the transformation law
for the functions $\sigma(t,\theta,\rho)$ and $A_i(t,\theta,\rho)$
but we will not give the explicit form here. Also, by acting on
the $G_{ij}$ part of the metric the diffeomorphism provides the
transformation law for $g_{ij}(t,\theta,\rho)$, namely
\be
&&{\cal L}_\xi g_{ij}=\alpha(\theta,\rho) (\partial_t+{2\over t}g_{ij})+\xi^\rho(\partial_\rho g_{ij}-{1\over \rho}g_{ij})\nonumber \\
&&+{\rho\over t}(A_i\partial_j\xi^\rho+A_j\partial_i\xi^\rho) +\nabla^{(g)}_i\xi_j+
\nabla^{(g)}_j\xi_i~~,
\lb{a11}
\ee
where $\xi^\rho$ takes the form (\ref{a6}). The transformations
generated by the $t$-independent part of the diffeomorphisms are
given in (\ref{sigma}), (\ref{Ai}) and (\ref{gij}) in the main
text.

It is interesting to note that the BMS transformation are a
particular case of the $t$-dependent diffeomorphisms.
Specifically, they are determined by the function $\alpha
(\theta,\rho)=f(\theta)\rho^{1/2}$ where $f(\theta)$ is an
arbitrary function of the angle coordinate  on the boundary
$\Sigma$ of the light-cone. The vector generating these
diffeomorphisms is
\be
\xi^t=f(\theta )\rho^{1/2} ~~, \nonumber \\
\xi^\rho=-{2\over t}f(\theta) \rho^{3/2}~~, \nonumber \\
\xi^i={1\over t}\rho^{3/2}~g_{(0)}^{ij}\partial_j f(\theta)~~.
\label{BMS}
\ee
The BMS transformations are known to be the transformations at null infinity
of asymptotically Minkowski spacetime (with a suitable definition
of the notion ``asymptotically Minkowski.'') It should be stressed
that our conditions for the asymptotic metric (\ref{6.1}) are more
general than that of BMS and thus the asymptotic diffeomorphisms
contain the BMS group as a subgroup.


\newpage

\begin{thebibliography} \\
\bibitem{tHooft} G. 't Hooft, ``Dimensional reduction in Quantum Gravity'', in
Salamfestschrift: A Collection of Talks, World Scientific Series in 20th
Century Physics, Vol. 4, ed. A. Ali, J. Ellis and S. Randjbar-Daemi (World
Scientific, 1993),
gr-qc/9310026.

\bibitem{Susskind} L. Susskind, ``The World as a Hologram'',
J. Math. Phys. {\bf 36} (1995) 6377, hep-th/9409089.

\bibitem{Malda} J. Maldacena, ``The Large N Limit of Superconformal
Field Theories and Supergravity'',
Adv. Theor. Math. Phys. {\bf 2} (1998) 231,
hep-th/9711200.

\bibitem{Gubs} S. Gubser, I. Klebanov and A. Polyakov,
``Gauge Theory Correlators from Non-Critical String Theory'',
Phys. Lett. {\bf B428} (1998) 105,
hep-th/9802109.


\bibitem{Wit} E. Witten, ``Anti De Sitter Space And Holography'',
Adv. Theor. Math. Phys. {\bf 2} (1998) 253, hep-th/9802150.


\bibitem{Wit1} E. Witten, ``Quantum Gravity In De Sitter Space'',
hep-th/0106109.

\bibitem{Str}  A. Strominger, ``The dS/CFT Correspondence'',
hep-th/0106113.

\bibitem{enbound}
R.~Bousso,
``A Covariant Entropy Conjecture,''
JHEP {\bf 9907}, 004 (1999)
[arXiv:hep-th/9905177].


\bibitem{polchinski}
J.~Polchinski,
``S-matrices from AdS spacetime,''
arXiv:hep-th/9901076.

\bibitem{Brown} J. D. Brown and M. Henneaux,
``Central charges in the canonical realization of asymptotic symmetries:
an example from three dimensional gravity'', Commun. Math. Phys. {\bf 104}
 (1986) 207.



\bibitem{Bondi:1962px}
H.~Bondi, M.~G.~van der Burg and A.~W.~Metner,
Waves From Axisymmetric Isolated Systems,''
Proc.\ Roy.\ Soc.\ Lond.\ A {\bf 269} (1962) 21.

\bibitem{Sommers}
P.~Sommers,
J.\ Math.\ Phys.\ {\bf 19} (1978), 549.

\bibitem{Ashtekar}
A.~Ashtekar, ``Asymptotic structure of the gravitational field at
spatial infinity'', In: {\it General Relativity and Gravitation}, ed. A.~Held
(Plenum Press, NY, 1980).

\bibitem{HE} S. W. Hawking and G. F. R. Ellis,
{\it The Large Scale Structure of Space-Time}
(Cambridge: Cambridge University Press).

\bibitem{FG} C. Fefferman and C. R. Graham: ``Conformal Invariants''. In:
{\it Elie Cartan et les Mathematiques d'aujord'hui}, (Asterisque, 1985),
95.

\bibitem{Shilov} I. M. Gel'fand and G. E. Shilov, {\it Generalized functions},
vol. I (Academic Press, NY, 1964).

\bibitem{Balasubramanian:2002zh}
V.~Balasubramanian, J.~de Boer and D.~Minic, ``Exploring de Sitter
space and holography,'' arXiv:hep-th/0207245.

\bibitem{Bertola}
J.~Bros, J.-P.~Gazeau  and U.~Moschella, ``Quantum Field Theory In
The De Sitter Universe,'' Phys.\ Rev.\ Lett.\ {\bf 73} (1994)
1746; J.~Bros and U.~Moschella,
Rev.\ Math.\ Phys.\  {\bf 8}, 327 (1996) [arXiv:gr-qc/9511019];
J.~Bros, H.~Epstein and U.~Moschella,
Commun.\ Math.\ Phys.\  {\bf 196}, 535 (1998)
[arXiv:gr-qc/9801099]; M.~Bertola, J.~Bros, V.~Gorini,
U.~Moschella and R.~Schaeffer,
Nucl.\ Phys.\ B {\bf 581}, 575 (2000) [arXiv:hep-th/0003098].

\bibitem{Bertola2}
M.~Bertola, V.~Gorini, U.~Moschella and R.~Schaeffer,
Phys.\ Lett.\ B {\bf 462}, 249 (1999) [arXiv:hep-th/9906035].

\bibitem{deHaro:2000wj}
S.~de Haro, K.~Skenderis and S.~N.~Solodukhin,
``Gravity in warped compactifications and the holographic stress tensor,''
Class.\ Quant.\ Grav.\  {\bf 18}, 3171 (2001)
[arXiv:hep-th/0011230].

\bibitem{GR} I. S. Gradshteyn and I. M. Ryzhik, {\it Tables of Integrals,
Series and Products} (New York: Academic Press).


\bibitem{st2}
R.~Bousso, A.~Maloney and A.~Strominger,
Phys.\ Rev.\ D {\bf 65}, 104039 (2002) [arXiv:hep-th/0112218].

\bibitem{Spradlin:2001nb}
M.~Spradlin and A.~Volovich, ``Vacuum states and the S-matrix in
dS/CFT,'' Phys.\ Rev.\ D {\bf 65}, 104037 (2002)
[arXiv:hep-th/0112223].


\bibitem{Balasubramanian:1999ri}
V.~Balasubramanian, S.~B.~Giddings and A.~E.~Lawrence,
JHEP {\bf 9903}, 001 (1999)
[arXiv:hep-th/9902052].


\bibitem{Giddings:1999qu}
S.~B.~Giddings,
Phys.\ Rev.\ Lett.\  {\bf 83}, 2707 (1999)
[arXiv:hep-th/9903048].



\bibitem{BjorkenDrell} J.~D. Bjorken and S.~D.~Drell, {\it Relativistic Quantum fields},
vol.I (McGRaw-Hill, Inc.).

\bibitem{HS} M. Henningson and K. Skenderis, ``The holographic Weyl
anomaly'', JHEP {\bf 9807} (1998), 023,
hep-th/9806087; ``Holography and the Weyl anomaly'', hep-th/9812023.


\bibitem{ISTY} C. Imbimbo, A. Schwimmer, S. Theisen and S. Yankielowicz,
``Diffeomorphisms and Holographic
Anomalies'', Class. Quant. Grav. {\bf 17} (1999) 1129, hep-th/9910267.

\bibitem{mellin} F. Oberhettinger, {\it Tables of Mellin
Transforms,} Springer Verlag, New York, 1974.

\bibitem{SS} S.~N. Solodukhin, ``How to make the gravitational action on
non-compact space finite'', Phys.Rev. {\bf D62} (2000), 044016,
hep-th/9909197.

\bibitem{BS} R. Beig and B.G. Schmidt, ``Einstein's equations
near Spatial Infinity'', Commun.Math.Phys. {\bf 87} (1982) 65-80.

\bibitem{B} R. Beig, ``Integration of Einstein's equations near spatial
infinity'', Proc.R.Soc.Lond. {bf A 391} (1984) 295-304.



\bibitem{KS} K. Skenderis and S.~N. Solodukhin,
``Quantum effective action from the AdS/CFT correspondence'',
Phys. Lett. {\bf B432} (2000) 316-322, hep-th/9910023.


\bibitem{BrownYork} J.D. Brown and J.W. York,
``Quasilocal energy and conserved charges derived from the
gravitational action'',
Phys.Rev. {\bf D47} (1993), 1407-1419.




\end{thebibliography}
\end{document}